\documentclass[aps,prb,amsmath,amssymb,superscriptaddress,floatfix,showpacs,twocolumn,longbibliography]{revtex4-2}

\usepackage[T1]{fontenc}

\usepackage{graphicx}
\usepackage{mathdots}
\usepackage{epstopdf} 
\usepackage[usenames,dvipsnames]{color}
\usepackage{bm}

\usepackage{inputenc}
\usepackage{xcolor}
\usepackage{tikz}
\usepackage{physics}

\usepackage{booktabs}

\usepackage[colorlinks,bookmarks=false,citecolor=blue,linkcolor=red,urlcolor=blue]{hyperref}
\usepackage[colorlinks,bookmarks=false,citecolor=blue,linkcolor=red,urlcolor=blue]{hyperref}
\usepackage{xcolor}
\usepackage{mathdots}
\usepackage{float}
\usepackage{needspace}
\usepackage{soul}
\usepackage[usenames,dvipsnames]{color}

\newcommand{\noteZ}[1]{{\color{blue} \emph{ GyZ: #1}}}

\begin{document}

\preprint{APS/123-QED}

\title{Limitations of the $g$-tensor formalism of semiconductor spin qubits}

\author{Zolt\'an György}
\affiliation{ELTE Eötvös Loránd University, Institute of Physics, H-1117 Budapest, Hungary}

\author{Andr\'as P\'alyi}
\affiliation{Department of Theoretical Physics, Institute of Physics, Budapest University of Technology and Economics, M\H{u}egyetem rkp. 3., H-1111 Budapest, Hungary}
\affiliation{HUN-REN-BME-BCE Quantum Technology Research Group, Műegyetem rkp. 3., H-1111 Budapest, Hungary}

\author{G\'abor Sz\'echenyi}
\affiliation{ELTE Eötvös Loránd University, Institute of Physics, H-1117 Budapest, Hungary}

\date{\today}

\begin{abstract}
The $g$-tensor formalism is a powerful method for describing the electrical driving of semiconductor spin qubits. However, up to now, this technique has only been applied to the simplest qubit dynamics, resonant monochromatic driving by a single gate. Here we study the description of (i) monochromatic driving using two driving gates and bichromatic driving via (ii) one or (iii) two gates. Assuming a general Hamiltonian with qubit states well separated from excited orbital states, we find that when (i) two driving gates are used for monochromatic driving or (ii) a single one for bichromatic, the $g$-tensor formalism successfully captures the leading-order dynamics. We express the Rabi frequency and the Bloch-Siegert shift using the $g$-tensor and its first and second derivatives with respect to the gate voltage. However, when (iii) bichromatic driving is realized using two distinct driving gates, we see a breakdown of $g$-tensor formalism: the Rabi frequency cannot be expressed using the $g$-tensor and its derivatives. We find that beyond the $g$-tensor and its derivatives, three additional parameters are needed to capture the dynamics. We demonstrate our general results by assuming an electron (hole) confined in a circular quantum dot, subjected to Rashba spin-orbit interaction. 
\end{abstract}

\maketitle

\section{Introduction}
\label{introduction}

The spin of an electron or hole confined within a semiconductor quantum dot is a promising candidate for qubit implementation \cite{loss1998quantum,burkard1999coupled, fang2023recent,burkard2023semiconductor}, due to the small footprint and potential in scalability \cite{li2018crossbar, philips2022universal, hendrickx2021four,borsoi2024shared,zhang2024universal,wang2024operating, john2024two,george202412, lim20242x2}, high fidelity single \cite{veldhorst2014addressable,yoneda2018quantum,lawrie2023simultaneous,mills2022high,huang2024high} and two-qubit operations \cite{xue2022quantum,mills2022two,noiri2022fast,huang2024high}, and the compatibility with quantum error correction schemes \cite{takeda2022quantum,van2022phase,hetenyi2024tailoring,pataki2024compiling}. In spin qubit experiments, an external magnetic field is usually applied to lift the Kramers degeneracy of the states. The interaction between the magnetic field and the spin in the presence of spin-orbit interaction (SOI) is effectively described by the $g$-tensor $\hat{g}(V_g)$, which depends on the static gate voltages $V_g$ applied on the surrounding metallic electrodes \cite{venitucci2018electrical,crippa2018electrical,liles2021electrical,hendrickx2024sweet}.

\begin{figure}[h!]
\centering
\includegraphics[width=\columnwidth]{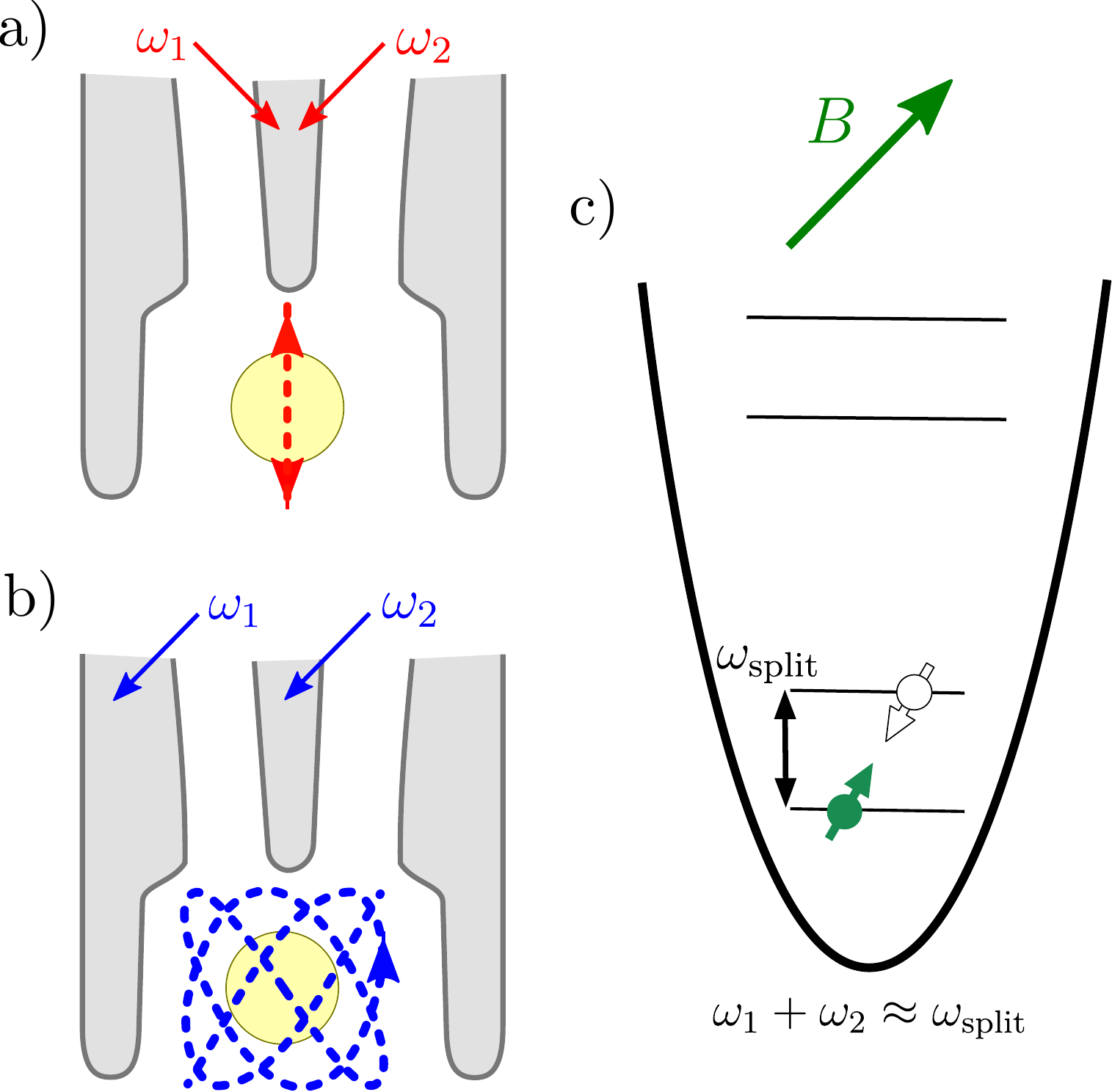}

\caption {\label{fig:system} \footnotesize{\textbf{Bichromatic driving of the qubit.} a) Two signals of different frequencies ($\omega_1$ and $\omega_2$) modulate the qubit, both applied to the same gate. As a result, the quantum dot (yellow dot) moves back and forth (red dashed lines). b) If the signals are applied to different gates, then the quantum dot moves on a Lissajous-like curve (blue dashed line). 
c) The qubit is defined by the two spin states of the lowest orbital level split by the external magnetic field $B$. The qubit is driven in a bichromatic way if the resonance condition $\omega_1+\omega_2\approx \omega_\mathrm{split}$ is fulfilled.}}
\end{figure}
The substantial spin-orbit interaction present in semiconductor materials \cite{rashba2003orbital,golovach2006electric,nowack2007coherent,watzinger2018germanium,hendrickx2020single} or magnetic field gradients provided by micromagnets \cite{tokura2006coherent,pioro2008electrically} enable the implementation of electric-dipole spin resonance (EDSR). Usually, coherent spin manipulation is achieved by applying ac voltages to the electrodes with a microwave frequency resonant with the qubit's energy splitting, however recently an operation scheme based on hopping spins has been demonstrated using both germanium and silicon spin qubits \cite{wang2024operating, unseld2024baseband}. Theoretical and experimental studies describe the resonant driving by the $g$-tensor formalism ($g$-TF) \cite{venitucci2018electrical,crippa2018electrical}, meaning that in the Zeeman spin Hamiltonian the static gate-voltage dependence of the $g$-matrix is replaced by its dependence on the ac voltage, $\hat{g}(V_g) \rightarrow \hat{g}(V_g^{(0)}+V_{ac}\cos{\omega t})$. As a result,  key quantities such as the frequency of the coherent oscillation, the so-called Rabi frequency, can be easily derived. Although this theoretical approach to describe coherent control is straightforward and intuitive, it is heuristic and hence has to be benchmarked against systematic perturbative or numerically exact techniques.

Two distinct EDSR mechanisms are commonly identified in the literature: $g$-tensor magnetic resonance and iso-Zeeman EDSR \cite{crippa2018electrical}. In the former, an ac field modulates the confinement potential, which leads to a time-dependent $g$-tensor of the electron (hole). In the latter mechanism, while the $g$-tensor remains time-independent and spatially uniform, an effective ac magnetic field emerges due to the interplay between spin-orbit coupling and the ac electric field. The two mechanisms can be distinguished through the behavior of the Zeeman tensor $^t\hat{g}\cdot\hat{g}$: in the $g$-tensor magnetic resonance, the tensor has time dependence, whereas in the iso-Zeeman EDSR, it remains constant over time. It has been shown in the literature that in the case of monochromatic driving using a single driving gate both mechanisms can be handled within a unified framework using the $g$-TF \cite{crippa2018electrical}. However, whether the $g$-TF can describe a more general qubit dynamics has not been investigated.

One of the pressing challenges in semiconductor-based quantum technology is scaling up the number of qubits. Recent progress has led to the development of increasingly large 2D spin-qubit architectures \cite{hendrickx2021four,wang2024operating}. However, a major hurdle in these systems is how to control individual qubits efficiently while keeping the number of control electrodes to a minimum \cite{,li2018crossbar,franke2019rent,boter2022spiderweb,paraskevopoulos2023spinq,borsoi2024shared}. One promising approach is using crossbar architecture combined with bichromatic electric control, where two distinct ac voltages are applied to the gate electrodes \cite{gyorgy2022electrically}. By controlling the sum or difference of their frequencies to match the qubit splitting, coherent Rabi oscillations can be realized. Bichromatic operations have been successfully demonstrated recently using germanium hole qubits \cite{john2024bichromatic}.

A key question arises whether the $ g$-TF can accurately describe qubit dynamics in the case of monochromatic driving using two gates or bichromatic driving.
In this paper, we compare results for the Rabi frequency and Bloch-Siegert shift obtained from $g$-TF against an analytical perturbative technique (time-dependent Schrieffer-Wolff transformation, TDSW) and the numerically exact solution of the time-dependent Schrödinger-equation. We show that when two ac signals are multiplexed onto the same gate electrode, then the $g$-TF successfully captures the qubit dynamics. However, when the ac signals are applied to different gate electrodes, an additional term appears in the effective Hamiltonian that cannot be derived from the $g$-tensor. To demonstrate our general findings, we theoretically examine the bichromatic excitation of an electron in a circular quantum dot with Rashba SOI (See Fig. \ref{fig:system}). 

The structure of the paper is as follows: In Section \ref{gTMR},  we review known results from the literature, including the derivation of the gate-voltage-dependent $g$-tensor and the application of $g$-TF for standard (monochromatic) EDSR. In Section \ref{gTMR_mono_two} we present $g$-TF for monochromatic driving using two different gates, while in Section \ref{gTMR_bichro}, we derive the effective Hamiltonian for a qubit under bichromatic excitation and analyze the conditions for the validity of $g$-TF. Section \ref{Rashba_model} presents a detailed case study of bichromatic control of a circular quantum dot in the presence of Rashba SOI. We also give perturbative expressions for the Rabi frequency and for the deviation from resonance, known as the Bloch-Siegert shift. The paper concludes with a discussion and a conclusion.

\section{$g$-tensor formalism}
\label{gTMR}

In this section, we briefly review the derivation of the gate-voltage-dependent $g$-tensor from a multilevel quantum dot Hamiltonian. In Ref. \onlinecite{venitucci2018electrical} it is shown that in the presence of a modulated gate voltage, the time-dependent Hamiltonian governing the effective monochromatic dynamics of qubits can be accurately derived by using the $g$-TF.  Consequently, the Rabi frequency can be directly calculated from the gate-voltage dependent $g$-tensor.

Let us consider a single quantum dot with an electron (hole) defined by the Hamiltonian
\begin{equation}
\label{eq:H}
H = H_\textrm{kin} + H_\textrm{conf}(V_g) + H_\textrm{SO} + H_B,
\end{equation}
where the terms describe the kinetic energy, confinement potential, spin-orbit interaction, and the Zeeman effect of the static magnetic field $\bm{B}$, respectively. We denote that the shape of the confinement potential depends on the voltage $V_g$ of a gate measured from an offset value $V_g^{(0)}$. Assuming weak magnetic fields and small gate voltages,  we treat 
\begin{equation}
\label{eq:Hunperturb}
H^{(0)} = H_\textrm{kin} + H_\textrm{conf}(V_g^{(0)}) + H_\textrm{SO}
\end{equation}
as the unperturbed Hamiltonian, while the remaining terms $H_P=H-H^{(0)}$ are considered as the perturbation. 

All terms in the unperturbed Hamiltonian, $H^{(0)}$, preserve time-reversal symmetry, which leads to doubly degenerate energy levels, commonly known as Kramers doublets.  The perturbative Hamiltonian, $H_P$, typically couples the ground-state and excited-state subspaces, however, an appropriately chosen unitary transformation (Schrieffer–Wolff transformation) decouples the ground-states from the high-energy subspaces. This procedure results in a $2\times 2$ effective Hamiltonian that governs the dynamics within the perturbed ground-state subspace.  The effective Hamiltonian, after it is linearized with respect to  the magnetic field, has the form of
\begin{equation}
\label{eq:H_eff}
H_\textrm{eff}=\frac{1}{2} \mu_B \boldsymbol{\sigma}\cdot \hat{g}(V_g) \boldsymbol{B},
\end{equation}
where $\mu_B$ is the Bohr-magneton, $\boldsymbol{\sigma} =\left\{\sigma_x ,\sigma_y, \sigma_z \right\}$ are Pauli matrices acting in the perturbed ground-state subspace, and $\hat{g}(V_g)$ is the gate-voltage dependent $g$-tensor represented by a $3\times 3$ real matrix.  We omit the magnetic-field-independent term of the effective Hamiltonian because it describes a collective energy shift in the ground-state subspace. According to the effective Hamiltonian,  in an external magnetic field the Kramers degeneracy of the ground states is lifted, and the splitting energy is  $\hbar\omega_\textrm{split}= \mu_B\left|\hat{g}(V_g) \boldsymbol{B}\right|$.   In this system,  we define our qubit states as the ground and excited states of Eq.~\eqref{eq:H_eff}.

Suppose that the gate voltage is not static but modulated in time $V_g(t)=V_g^{(0)}+V_\textrm{ac}\cos{\omega t}$. In that case,
the time-dependent effective Hamiltonian can be derived analogously to the method outlined in the previous paragraph. The key distinction lies in the time dependence of the perturbation, necessitating the use of the time-dependent Schrieffer-Wolff transformation \cite{schrieffer1966relation,bravyi2011schrieffer,romhanyi2015subharmonic}.
Namely, due to the time-dependent unitary transformation $U$, the Hamiltonian transforms as follows 
\begin{equation} \label{eq:UHU}
 \Tilde{H}(t)=U(t)HU^\dag(t) +i \hbar \dot{U}(t) U^\dag(t),
 \end{equation}
resulting in an additional term $i\hbar \dot{U}(t) U^\dag(t)$ not present in the time-independent case. Subsequently,  it is necessary to determine whether this additional term provides a leading-order contribution (second order for monochromatic driving) to the effective Hamiltonian or not (for details, see App.~\ref{appendix_time_SW}). 
Specifically in the case of monochromatic driving, if $\hbar\omega$ is comparable to or smaller than the energy splitting $\hbar\omega_\textrm{split}$, the $i\hbar \dot{U} U^\dag$ term in leading order does not give a contribution to the  effective Hamiltonian:
\begin{equation}
\label{eq:H_efft}
H_\textrm{eff}(t)=\frac{1}{2} \mu_B \boldsymbol{\sigma}\cdot \hat{g}\left(V_g(t)\right) \boldsymbol{B}.
\end{equation}
We can see that it is not required to rederive the effective Hamiltonian in the case of time-dependent perturbation. Instead, one can substitute the dependence of the $g$-tensor on the static gate voltage in Eq.~\eqref{eq:H_eff}  with its dependence on the modulated gate voltage 
(for the proof, see App.~\ref{appendix_mono}). We refer to this procedure as $g$-TF.

When the modulation frequency $\omega$ equals $\omega_\textrm{split}$, coherent Rabi oscillation arises between the qubit states. The dynamics of the system is described by Eq.~\eqref{eq:H_efft}, from which the Rabi frequency can be derived,
\begin{equation}\label{eq:Crippa}
    \bm{f}_\mathrm{Rabi}=\frac{\mu_B  V_\mathrm{ac}}{2h \left|\hat g(V_g^{(0)})\mathbf{B}\right|}\left[\hat g(V_g^{(0)}) \mathbf{B} \right] \times \left[ \hat g'(V_g^{(0)})\mathbf{B} \right].
\end{equation}
This expression is identical to Eq. (3) of Ref. \onlinecite{crippa2018electrical}. In the case of monochromatic driving, according to the leading-order calculation in the Rabi frequency, the shift in the resonance condition, referred to as the Bloch-Siegert shift, $\omega_\textrm{BS}=\omega-\omega_\textrm{split}$ is found to be zero.

\section{$g$-tensor formalism for monochromatic driving using two gates}\label{gTMR_mono_two}
Our first step in studying the validity of $g$-TF is considering monochromatic driving using two different gates, which can be plunger or barrier gates for instance. In this case, the confinement potential depends on both gate voltages; therefore, we explicitly denote the dependence of the $g$-tensor $\hat{g}(V_{g1},V_{g2})$ on both gate voltages, $V_{g1}$ and $V_{g2}$. 

When gate voltages are modulated $V_{g1(2)}(t)=V_{g1(2)}^{(0)} + \delta V_{1(2)}(t)$,  the following effective time-dependent Hamiltonian can be derived based on the $g$-TF:
\begin{eqnarray} \label{eq:mono_twogate}
 H_\textrm{eff}(t)&=&\frac{1}{2}\mu_B\bm{\sigma}\cdot\left[\hat{g}(V_{g1}^{(0)},V_{g2}^{(0)})+\right.\nonumber\\ &+&\left.\frac{\partial \hat{g}}{\partial V_{g1}}\delta V_1(t)+\frac{\partial \hat{g}}{\partial{V}_{g2}}\delta V_2(t) \right]\bm{B}, 
\end{eqnarray}
where the partial derivatives are evaluated at $V_{g1}=V_{g1}^{(0)}$ and $V_{g2}=V_{g2}^{(0)}$.  We introduce the following vectors: 
\begin{eqnarray}
    &\hbar\bm{\Omega}=\mu_B\hat{g}(V_{g1}^{(0)},V_{g2}^{(0)}) \bm{B}, \hspace{3mm} \hbar\bm{\Omega}_1'=\frac{\mu_B}{2}\frac{\partial \hat{g}}{\partial V_{g1}} \bm{B}, \nonumber\\ & \hbar\bm{\Omega}_2'=\frac{\mu_B}{2}\frac{\partial \hat{g}}{\partial V_{g2}} \bm{B}.
\end{eqnarray}
Vectors $\bm{\Omega}_1'$ and $\bm{\Omega}_2'$ can be written as sum of components parallel and perpendicular to $\bm{\Omega}$, $\bm{\Omega}_{1(2)}'=\bm{\Omega}_{1(2)\parallel}'+\bm{\Omega}_{1(2)\perp}'$. We confirm the validity of the $g$-TF by comparing Eq.~\eqref{eq:mono_twogate} with the corresponding result obtained from second-order TDSW, as discussed in App.~\ref{appendix_mono}.

Let us assume sinusoidal gate modulations with initial phases $\phi_1$ and $\phi_2$, $\delta V_{1(2)}(t)=V_\mathrm{ac,1(2)}\cos{(\omega_{1(2)}t+\phi_{1(2)})}$. The Rabi frequency can be expressed as: 
\begin{equation}\label{eq:Rabi_mono_twogate}
    f_\mathrm{Rabi}=\frac{1}{2\pi}\sqrt{f_\mathrm{1}^2+f_\mathrm{2}^2+2f_\mathrm{1}f_\mathrm{2}\cos{(\chi_2-\chi_1+\phi_2-\phi_1)}},
\end{equation}
where $f_{1(2)}=\lvert\bm{\Omega}_{1(2)\perp}'\rvert V_\mathrm{ac,1(2)}$ and $\chi_2-\chi_1$ is the azimuthal angle of $\bm{\Omega}'_{2\perp}$ in the Cartesian coordinate system spanned by $\bm{\Omega}'_{1\perp}$, $\bm{\Omega}\times \bm{\Omega}'_{1\perp}$ and $\bm{\Omega}$. If the strengths of the two modulations are chosen such that $f_1=f_2$, the Rabi frequency can be set to zero by using an appropriate phase difference, as shown for specific models in Ref. \onlinecite{dey2024role}. This setting corresponds to the special case of electron spin resonance when the drive is circularly polarized and counter-rotating.

\section{$g$-tensor formalism for bichromatic driving}
\label{gTMR_bichro}

In Section \ref{gTMR}, we reviewed the validity of $g$-TF for monochromatic driving using a single gate, in Section \ref{gTMR_mono_two} we presented $g$-TF for monochromatic driving using two gates. However, the applicability of $g$-TF to describe more complex dynamics 
than monochromatic driving remains unaddressed in the existing literature. In this chapter, we specifically investigate this topic in the context of bichromatic driving, where the spin is driven by two ac fields with distinct frequencies, the sum of which equals the spin qubit's splitting.
We investigate two distinct realizations of bichromatic driving: (i) multiplexing both ac signals onto a single gate (Fig. \ref{fig:system}a), and (ii) applying each ac signal to separate gates (Fig. \ref{fig:system}b). Our main finding is that in the latter case, the effective Hamiltonian constructed from the heuristic $g$-TF is invalid: the leading-order Rabi frequency formula obtained from the $g$-TF effective Hamiltonian is different from that obtained from TDSW. We also note that contrary to monochromatic driving using two different gates (see Eq.~\eqref{eq:Rabi_mono_twogate}), the bichromatic Rabi frequency is unaffected by the initial phases of the driving signals (the initial phase difference will be lost during the time evolution due to the different driving frequencies).  

Similarly to Section \ref{gTMR}, we consider $H^{(0)}$ from Eq.~\eqref{eq:Hunperturb} as the unperturbed system, while the magnetic field and the modulation of gate voltages are treated as perturbations. Our calculations are carried out on the eigenbasis of the unperturbed Hamiltonian $H^{(0)}$
\begin{equation}\label{eq:basis}
    H^{(0)}\ket{\Psi_{k\alpha}}=E_k\ket{\Psi_{k\alpha}}, 
\end{equation}
where $k$ denotes the non-negative integer orbital index, and $\alpha$ labels the Kramers-degenerate pseudospin states. The form of the perturbation differs for the two distinct types of bichromatic driving, and thus, they are discussed separately in the upcoming two subsections. 

\subsection{Driving with a single gate}

In the case of bichromatic driving using a single gate, the gate voltage is modulated as $V_g(t)=V_g^{(0)}+\delta V(t)$, where
\begin{equation}\label{eq:modulation_single}
\delta V(t)=V_\mathrm{ac,1}\cos{\omega_1 t}+V_\mathrm{ac,2}\cos{\omega_2 t}.
\end{equation}
Hence, the perturbation has the form 
$H_P=H_\textrm{conf}(V_g(t))- H_\textrm{conf}(V_g^{(0)})+ H_B.$
To obtain the effective Hamiltonian $H_\mathrm{eff}(t)$ we use time-dependent Schrieffer-Wolff transformation on the Hamiltonian, projecting onto the two lowest-energy $k=0$ qubit states. A third-order TDSW has to be applied to capture the bichromatic dynamics, yielding the effective Hamiltonian: 
\begin{eqnarray} \label{eq:effective}
 H_\textrm{eff}(t)&=&\frac{1}{2}\mu_B\bm{\sigma}\cdot\left[\Hat{g}(V_g^{(0)})+\Hat{g}'(V_g^{(0)})\delta V(t)+\right.\nonumber\\ &+&\left.\frac{\Hat{g}''(V_g^{(0)})}{2}\delta V(t)^2 \right]\bm{B}, 
\end{eqnarray}
where the first and second derivatives of the $g$-tensors with respect to the static gate voltage appear. We observe that this result is equivalent to deriving the Hamiltonian $H_\textrm{eff}=\frac{1}{2}\mu_B\bm{\sigma}\cdot\Hat{g}(V_{g}(t))\bm{B}$ using the $g$-TF, followed by a second-order expansion of it in $V_g$ and a first-order in $B$. The $g$-TF provides a correct result for deriving the effective Hamiltonian, meaning that the second term in Eq.~\eqref{eq:UHU} does not contribute to the dynamics, as shown in Appendix \ref{App_general_oneop}.

As a consequence of the validity of the $g$-TF, the Rabi frequency and the Bloch-Siegert shift can be expressed in terms of the $g$-tensor and its derivatives, analogous to the monochromatic case described in Eq.~\eqref{eq:Crippa}. To derive these quantities, one must first construct the Floquet matrix from the Hamiltonian given in Eq.~\eqref{eq:effective}. Subsequently, after performing a (time-independent) Schrieffer-Wolff transformation on the Floquet matrix, we derive the Rabi frequency for the bichromatic driving
\begin{equation}\label{eq:gTMR_Rabi_bi}
    f_\mathrm{Rabi}=\left\lvert \bm{\Omega}''_\perp-\frac{\Omega\Omega'_\parallel }{\omega_1 \omega_2}\bm{\Omega}'_\perp\right\rvert \frac{V_\mathrm{ac,1}V_\mathrm{ac,2}}{2\pi}, 
\end{equation}
as well as the Bloch-Siegert shift 
\begin{equation}
\begin{aligned}
\label{eq:gTMR_BS}\omega_\mathrm{BS}&=\left(\Omega''_\parallel+\frac{\Omega\Omega_\perp^{\prime 2}}{\omega_2(\omega_2+2\omega_1)}\right)V_\mathrm{ac,1}^2+\\&+\left(\Omega''_\parallel+\frac{\Omega\Omega_\perp^{\prime 2}}{\omega_1(\omega_1+2\omega_2)}\right)V_\mathrm{ac,2}^2
\end{aligned}
\end{equation}
describing the deviation from the resonance condition
\begin{equation}
    \omega_1+\omega_2=\Omega+\omega_\mathrm{BS}. 
\end{equation}
A detailed derivation is provided in App.~\ref{App_general_oneop}. In these formulas we introduce $\hbar\bm{\Omega}=\mu_B\Hat{g}(V_g^{(0)}) \bm{B}$, $\hbar\bm{\Omega'}=\mu_B\Hat{g}'(V_g^{(0)})\bm{B}/2$ and $\hbar\bm{\Omega''}=\mu_B\Hat{g}''(V_g^{(0)})\bm{B}/4$. Furthermore, we denote the projections of vectors $\bm{\Omega'}$ and  $\bm{\Omega''}$ that are parallel and perpendicular to  $\bm{\Omega}$ as:  $\bm{\Omega'}_\perp$, $\bm{\Omega'}_\parallel$, $\bm{\Omega''}_\perp$, and $\bm{\Omega''}_\parallel$. For simplicity, the magnitudes of these vectors are represented by non-bold characters.

The Rabi frequency is given by a third-order equation in the perturbation, as expressed in Eq.~\eqref{eq:gTMR_Rabi_bi}, which includes the interference of two contributions containing the second derivative of the $g$-tensor or the first derivatives of the $g$-tensor. The Bloch-Siegert shift, given by a third-order term in the perturbation, also depends on both the first and second derivatives of the $g$-tensor.

It is worth noting that during the derivation of Eq.~\eqref{eq:gTMR_Rabi_bi} and Eq.~\eqref{eq:gTMR_BS} from the effective Hamiltonian Eq.~\eqref{eq:effective}, we imposed the condition that the strengths and frequencies of the two ac fields should be of the same order of magnitude, but  $\omega_1\neq \omega_2$. Furthermore, the parameters were required to satisfy the following hierarchy,
$
\Omega \gg \Omega' V_\mathrm{ac,1(2)} \gg  \Omega'' V_\mathrm{ac,1(2)}^2
$, which generally holds under weak driving, except for a few fine-tuned magnetic field orientations. 

\subsection{Driving with two gates}

When we describe qubit control using two different gates, it is necessary to account for the dependence of the confining potential and the static $g$-tensor on both gate voltages, $V_\textrm{conf}(V_{g1},V_{g2})$ and $\Hat{g}(V_{g1},V_{g2})$.  If the two ac fields are switched on, the gate voltages are modulated as, $V_{g1}(t)=V_{g1}^{(0)} + V_\mathrm{ac,1}\cos{\omega_1 t}$ and $V_{g2}(t)=V_{g2}^{(0)} + V_\mathrm{ac,2}\cos{\omega_2 t}$. We consider $H_P=H_\textrm{conf}(V_{g1}(t), V_{g2}(t))- H_\textrm{conf}(V_{g1}^{(0)}, V_{g2}^{(0)})+ H_B$  as a perturbation.

Similar to the previous subsection, we derive the effective Hamiltonian from the multi-level model using a third-order time-dependent Schrieffer-Wolff transformation,
\begin{equation} \label{eq:effective_two_punger}
H_\textrm{eff}(t)=H_\textrm{eff}^{g-\mathrm{TF}}(t) + H_\textrm{eff}^\textrm{TD}(t).
\end{equation}
In this derivation, we obtain $H_\textrm{eff}^{g-\mathrm{TF}}(t)$ the third-order expansion of the effective operator $H_\textrm{eff}=\frac{1}{2}\mu_B\bm{\sigma}\cdot\Hat{g}(V_{g1}(t),V_{g2}(t))\bm{B}$ derived from $g$-TF, as well as an additional term $H_\textrm{eff}^\textrm{TD}(t)$ of comparable magnitude arising due to the time-dependent nature of the Schrieffer-Wolff transformation.  It becomes clear that, in the case of bichromatic driving using two gates, the pure $g$-TF alone does not accurately describe the effective Hamiltonian.  The extra term cannot be expressed using the static $g$-tensor alone but must be instead represented as a sum over higher-lying energy states. The matrix elements of it are given as:
\begin{equation} \label{eq:sum_formula}
\left[H_\textrm{eff}^\textrm{TD}(t)\right]_{\alpha\beta}=\frac{i\hbar}{2}\sum\limits_{l,\delta}\frac{\dot{H}_{P,0\alpha l\delta}H_{P,l\delta 0\beta}-H_{P,0\alpha l\delta}\dot{H}_{P,l\delta 0\beta}}{(E_l-E_0)^2},
\end{equation}
where $l>0$ denotes the integer orbital indices,
furthermore,  $\alpha$, $\beta$ and $\delta$ label pseudospin states. The index zero represents the ground state. We use a compact form $H_{P,l\delta 0\beta}$ instead of $\langle \Psi_{l\delta} | H_{P}| \Psi_{0\beta}\rangle$. The dot represents the time derivative. In Eq. (\ref{eq:sum_formula}) to capture the bichromatic dynamics in leading order, only the first-order expansion of the confinement modulation needs to be considered.  In the case of bichromatic driving using a single gate, Eq. (\ref{eq:sum_formula}) does not contribute to the dynamics, for more details see App.~\ref{App_general_oneop}.

The time-dependent Hamiltonian in  Eq.~\eqref{eq:sum_formula} contains multiple magnetic and electric terms, among which only the following contribute to the bichromatic driving under the resonance condition $\omega_1+\omega_2=\omega_\mathrm{split}+\omega_\mathrm{BS}$:  
\begin{equation}
    H_\mathrm{eff}^{\mathrm{TD}}(t)=V_\mathrm{ac,1}V_\mathrm{ac,2}(\omega_1-\omega_2)\sin{[(\omega_1+\omega_2)t]}\bm{\Upsilon}
    \cdot\bm{\sigma},
\end{equation}
where $\bm{\Upsilon}$ is a three-component vector. Only the component perpendicular to $\bm{\Omega}$,  $\bm{\Upsilon_\perp}$ contributes to the bichromatic dynamics. Consequently, Eq.~\eqref{eq:sum_formula}, the contribution to the effective Hamiltonian beyond the $g$-tensor formalism $g$-TF, can be effectively described using only two parameters, for a given magnetic field direction. The expression for $\bm{\Upsilon}$, the Rabi frequency, and the Bloch-Siegert shift are provided in App.~\ref{App_bichro_two}. 

\section{Circular quantum dot with Rashba SOI}
\label{Rashba_model}
In this section, we investigate the validity of the $g$-TF  through a simple model: an electron confined in a two-dimensional parabolic quantum dot with Rashba spin-orbit interaction, the Hamiltonian:
\begin{equation}\label{eq:Rashba_Ham}
    H=H_0+H_\mathrm{SO}+H_B+H_E(E(t)).
\end{equation}
Here, $H_0$ is a Hamiltonian of a two-dimensional harmonic oscillator in the $x-y$ plane
\begin{equation}
    H_0=\frac{p_x^2+p_y^2}{2m}+\frac{m \omega_0^2}{2}(x^2+y^2), 
\end{equation}
where $m$ is the effective mass and $\omega_0$ is  the angular frequency of the oscillator. $H_B$ describes the interaction with an external in-plane magnetic field $\bm{B}$: 
\begin{equation}\label{eq:Zeeman}
    H_B=\frac{1}{2}g \mu_B \bm{B}\cdot \bm{\sigma}=\frac{1}{2}\Tilde{B}\bm{b}\cdot\bm{\sigma}, 
\end{equation}
where we introduce $\Tilde{B}=g\mu_B |\bm{B}|$, a quantity with energy dimension. The direction of the magnetic field is represented by the vector $\bm{b}=\left(\cos{\phi},\sin{\phi},0 \right)$. $H_\mathrm{SO}$ describes the Rashba spin-orbit interaction: 
\begin{equation}
    H_\mathrm{SO}=\alpha(p_x\sigma_y-p_y\sigma_x),  
\end{equation}
where $\alpha$ is the spin-orbit coupling strength. For convenience, we later use $\Tilde{\alpha}=\alpha\sqrt{m\hbar\omega_0/2}$, a quantity with energy dimension.

In contrast to the previous chapter, for simplicity, instead of the gate voltages, we describe the dynamics by the homogeneous and time-dependent electric field at the position of the qubit. In this case, by applying the $g$-TF, we can derive the effective Hamiltonian as follows.  First, consider the Hamiltonian in the presence of a constant electric field and treat $H_B$ and $H_E(E_\textrm{dc})$ as perturbations. Using the time-independent Schrieffer-Wolff transformation, we derive the effective Hamiltonian $H_\textrm{eff}(E_\textrm{dc})$. According to the $g$-TF,  in the case of a time-dependent electric field, we simply replace the time-independent variable in the effective Hamiltonian with the time-dependent one: $H_\textrm{eff}^{g-\textrm{TF}}(E(t))$. 

It is well known from the literature \cite{golovach2006electric, crippa2018electrical, venitucci2018electrical} that the $g$-TF provides an efficient description of Rabi dynamics in a circular quantum dot with Rashba spin-orbit interaction when it is driven by a single gate monochromatically. In the following, we investigate the model in the case of bichromatic excitations which can be realized either via multiplexing two ac signals on the same gate or by using two different driving gates.

\subsection{Bichromatic driving with a single gate}

Let us assume that the modulation of the gate voltage induces a homogeneous electric field in the $x$-direction at the position of the quantum dot.  If two ac signals with different frequencies are multiplexed on a gate electrode,  the effect of the resulting electric field can be described by the  Hamiltonian
\begin{equation}\label{eq:electric_oneop}
    H_E=e x (E_\mathrm{ac,1}\sin{\omega_1 t}+E_\mathrm{ac,2}\sin{\omega_2 t}),
\end{equation}
where $e$ is the elementary charge, and $E_\mathrm{ac,1}$ and $E_\mathrm{ac,2}$ represent the amplitudes of the electric fields. For subsequent use, we introduce two new quantities with energy dimension, $\Tilde{E}_\mathrm{ac,1}=eE_\mathrm{ac,1}\sqrt{\hbar/(2m\omega_0)}$ and $\Tilde{E}_\mathrm{ac,2}=eE_\mathrm{ac,2}\sqrt{\hbar/(2m\omega_0)}$.

For bichromatic qubit control, the two driving frequencies should be selected such that their sum closely matches the qubit's splitting frequency.  The frequency of the resulting Rabi oscillation is determined using three distinct approaches: (1) within the framework of the $g$-TF, (2) through an analytical perturbative calculation, and (3) by numerically solving the time-dependent Schrödinger equation.

(1) The $g$-TF approach involves the following steps. First, we numerically diagonalize the Hamiltonian $H_0+H_\mathrm{SO}$, considering the 10 lowest-energy Kramers doublets. The perturbation, $H_B+H_E$, is treated analytically and is expanded on the diagonalized basis. We then derive the effective qubit Hamiltonian using a third-order time-independent Schrieffer-Wolff transformation. Finally, the Rabi frequency is extracted using two-mode Floquet theory \cite{shirley1965solution, ho1983semiclassical}.

(2) To perform analytical perturbative calculations, we treat not only the magnetic and electric fields but also the spin-orbit interaction as a perturbation:
\begin{equation}
     \frac{\Tilde{E}_\mathrm{ac,1}}{\hbar\omega_0}\approx \frac{\Tilde{E}_\mathrm{ac,2}}{\hbar\omega_0}\approx\frac{\Tilde{B}}{\hbar\omega_0}\approx \frac{\Tilde{\alpha}}{\hbar\omega_0} \ll 1.
\end{equation}
Treating $H_0$ as the unperturbed Hamiltonian and $H_\mathrm{SO}+H_B+H_E$ as the perturbation, the effective Hamiltonian can be derived using a fifth-order time-dependent Schrieffer-Wolff transformation. This allows us to determine the qubit splitting up to the fifth order,
\begin{equation}\label{eq:split}
\omega_\mathrm{split}=\frac{\Tilde{B}}{\hbar}-\frac{2\Tilde{B}\Tilde{\alpha}^2}{\hbar^3\omega_0^2}+\frac{8\Tilde{B}\Tilde{\alpha}^4-2\Tilde{B}^3\Tilde{\alpha}^2}{\hbar^5 \omega_0^4}.
\end{equation}
Subsequently, applying two-mode Floquet theory on the effective Hamiltonian,  we derive the fifth-order Rabi frequency,
\begin{equation}\label{eq:oneop_Rabi}
hf_\mathrm{Rabi}=\frac{2\Tilde{B}\Tilde{E}_\mathrm{ac,1}\Tilde{E}_\mathrm{ac,2}\Tilde{\alpha}^2\sin{2\phi}}{\hbar^4\omega_0^4}.
\end{equation}

In addition to the Rabi frequency, the shift of the resonance condition, the fifth-order Bloch-Siegert shift can also be determined using two-mode Floquet theory:
\begin{equation}\label{eq:BS1main}
    \omega_\mathrm{BS}=\sum_{k \in\{1,2\}}\frac{4\Tilde{B}\Tilde{\alpha}^2\omega_k^2\cos^2{\phi}}{\hbar^3\omega_0^4(\Tilde{B}^2-\hbar^2\omega_k^2)}\Tilde{E}_\mathrm{ac,k}^2. 
\end{equation}
More details about the calculation
can be found in App.~\ref{App_oneop}.

(3) During the numerical simulation, the Rabi frequency is determined as follows. First, we diagonalize the Hamiltonian $H_0+ H_\mathrm{SO}+H_B$, while considering the 10 lowest Kramers doublets. The ground state is chosen for the initial state of the simulation. The energy difference between the ground and first excited states defines the qubit splitting, $\hbar\omega_\textrm{split}$. One of the ac signal frequencies, $\omega_1$ is fixed, while the other is chosen as $\omega_2=\omega_\mathrm{split}+\omega_\mathrm{BS}-\omega_1$, where the analytically derived result, i.e. Eq.~\eqref{eq:BS1main}, is used for the Bloch-Siegert shift. The initial state is then evolved numerically using the total time-dependent Hamiltonian. The population of the first excited state exhibits Rabi oscillations as a function of time with values between 0 and larger than 0.95. We fit the numerical data to the off-resonant Rabi oscillation formula, getting the oscillation frequency directly from the fit. 

\begin{figure}
\centering
\includegraphics[width=\columnwidth]{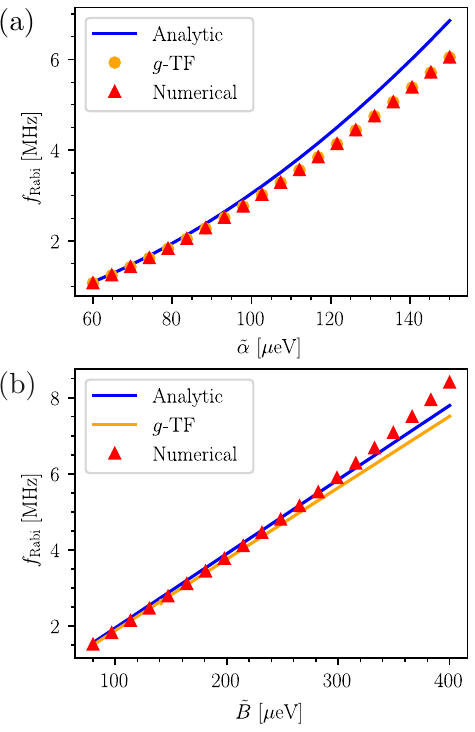}
\caption {\label{fig:oneop} \footnotesize{\textbf{Rabi frequencies of bichromatic driving with one gate.} The Rabi frequencies calculated using different methods are shown as a function of spin-orbit interaction strength $\Tilde{\alpha}$ (a) or magnetic field strength $\Tilde{B}$ (b). The blue curves show the result of the analytic calculation, see Eq.~\eqref{eq:oneop_Rabi}. The orange curves are the results of $g$-TF. The red curves are obtained by solving the Schrödinger-equation numerically. The parameters are $\hbar\omega_0=1000$ $\mu$eV, $\Tilde{B}=100$ $\mu$eV, $\Tilde{\alpha}=80$ $\mu$eV, $\Tilde{E}_\mathrm{ac,1}=70$ $\mu$eV, $\Tilde{E}_\mathrm{ac,2}=90$ $\mu$eV, $\phi=\pi/4$, $\omega_1=0.7 \Tilde{B}$, and $\omega_2$ was chosen according to the resonance condition.}}
\end{figure}

The Rabi frequencies obtained from the three different methods are plotted in Fig. \ref{fig:oneop} as a function of the spin-orbit coupling strength and the external magnetic field. It can be observed that for small values of $\Tilde{\alpha}$ and $\Tilde{B}$, the curves overlap, indicating that in the case of bichromatic driving using a single gate, the $g$-TF provides an accurate description. Another observation is that consistent with the analytical results, the Rabi frequency has a linear dependence on the external magnetic field and a parabolic dependence on the spin-orbit interaction strength.  As shown in Fig. \ref{fig:oneop}(a), for larger values of $\Tilde{\alpha}$, deviations between the different methods become apparent. This discrepancy arises because the numerical and $g$-TF methods treat the spin-orbit interaction exactly, while the analytical approach treats it only perturbatively. Similarly, for strong magnetic fields, slight differences between the results can be observed.

\subsection{Bichromatic driving using two gates}

In this subsection, we consider the control of a qubit using two distinct electrodes. The modulation of the voltage applied to each electrode generates a homogeneous electric field of different strength and direction at the location of the quantum dot. For simplicity, we assume that a voltage applied to one electrode induces an electric field in the $x$-direction, while a voltage applied to the other electrode generates a field in the $y$-direction. The corresponding Hamiltonian is given by
\begin{equation}\label{eq:electric_twoop}
           H_E=e x E_\mathrm{ac,1}\sin{\omega_1 t}+e y E_\mathrm{ac,2}\sin{\omega_2 t}.
\end{equation}

The Rabi frequency of bichromatic driving can be determined using three different approaches, as outlined in the previous subsection. The calculations follow the same procedure as described earlier. In the analytical perturbative calculation, we derive expressions for the Rabi frequency 
\begin{equation}\label{eq:twoop_Rabi}
    hf_\mathrm{Rabi}=\frac{4\Tilde{E}_\mathrm{ac,1}\Tilde{E}_\mathrm{ac,2}\Tilde{\alpha}^2\lvert \hbar\omega_1-\Tilde{B}\sin^2{\phi} \rvert}{\hbar^4 \omega_0^4}, 
\end{equation}
and for the Bloch-Siegert shift
\begin{equation}\label{eq:BS2main}
    \omega_\mathrm{BS}=\frac{4\Tilde{B}\Tilde{\alpha}^2\omega_1^2\cos^2{\phi}}{\hbar^3\omega_0^4(\Tilde{B}^2-\hbar^2\omega_1^2)}\Tilde{E}_\mathrm{ac,1}^2+\frac{4\Tilde{B}\Tilde{\alpha}^2\omega_2^2\sin^2{\phi}}{\hbar^3\omega_0^4(\Tilde{B}^2-\hbar^2\omega_2^2)}\Tilde{E}_\mathrm{ac,2}^2. 
\end{equation}

\begin{figure}
\centering
\includegraphics[width=\columnwidth]{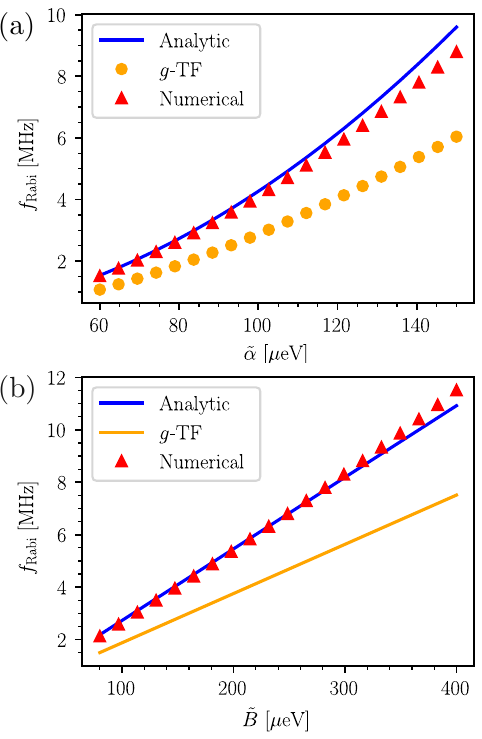}

\caption {\label{fig:twoop} \footnotesize{\textbf{Rabi frequencies of bichromatic driving with two gates.} The Rabi frequencies calculated using different methods are shown as a function of spin-orbit interaction strength $\Tilde{\alpha}$ (a) or magnetic field strength $\Tilde{B}$ (b). The blue curves show the result of the analytic calculation, see Eq.~\eqref{eq:twoop_Rabi}. The orange curves are the results of $g$-TF.  The red curves are obtained by solving the Schrödinger-equation numerically.  The parameters are $\hbar\omega_0=1000$ $\mu$eV, $\Tilde{B}=100$ $\mu$eV, $\Tilde{\alpha}=80$ $\mu$eV, $\Tilde{E}_\mathrm{ac,1}=70$ $\mu$eV, $\Tilde{E}_\mathrm{ac,2}=90$ $\mu$eV, $\phi=0$, $\omega_1=0.7 \Tilde{B}$, and $\omega_2$ was chosen according to the resonance condition.}}
\end{figure}

To compare the three methods, in Fig. \ref{fig:twoop} we plot the Rabi frequency as a function of both the spin-orbit interaction strength and the magnetic field. It is apparent that the results obtained using the $g$-TF method deviate from those of the other approaches. This discrepancy is a direct consequence of the general principle previously established, namely that the $g$-TF method is not applicable in the case of bichromatic driving with two gates, as it yields incorrect results. Conversely, the numerical and analytical methods provide accurate results. The corresponding curves exhibit good agreement for small values of $\Tilde{\alpha}$ and $\Tilde{B}$, diverging only at higher, non-perturbative values. As in the previous subsection, the Rabi frequency exhibits a linear dependence on the magnetic field and a parabolic dependence on the spin-orbit interaction strength.

\section{Discussion}
\label{discussion}
\subsection{Spin dynamics at zero magnetic field}
In Section \ref{gTMR_mono_two} of this work, we described the case of monochromatic driving with two gates, at finite magnetic field and resonant driving. We obtained the Rabi frequency (Eq. \eqref{eq:Rabi_mono_twogate}) using the $g$-TF via a second-order perturbative calculation. In the limit of zero magnetic field, the Rabi frequency of Eq. \eqref{eq:Rabi_mono_twogate} converges to zero. Apparently, this indicates that there is no spin dynamics in this zero-field limit. It is known that this conclusion is false: even at zero magnetic field, spin dynamics is induced by the spin-orbit-induced non-Abelian Berry phase, enabling holonomic gates \cite{san2008geometric,kolok2024protocols}. This dynamics is not described by the $g$-TF, since the effective Hamiltonian of Eq. \eqref{eq:mono_twogate}, obtained from the $g$-TF, vanishes at zero magnetic field. To study this zero-field dynamics, a third-order perturbative calculation needs to be carried out, incorporating the contribution of Eq. \eqref{eq:sum_formula} in the effective Hamiltonian. Compared to the second-order contribution of resonant driving from Eq. \eqref{eq:Rabi_mono_twogate}, this effect appears only as a correction to the result of the $g$-TF. It is beyond the scope of this work to study the zero-field dynamics in detail.

\subsection{Bichromatic driving with difference of the frequencies}
Throughout the paper, we considered the sum-frequency resonance caused by bichromatic driving, i.e., the resonance when $\omega_1+\omega_2=\omega_\mathrm{split}+\omega_\mathrm{BS}$. Bichromatic driving can induce a difference-frequency resonance as well, when $\omega_1-\omega_2=\omega_\mathrm{split}+\omega_\mathrm{BS}$. An advantage of the sum-frequency resonance is that it requires lower microwave frequencies, might be easier to generate and deliver to the gates. Another advantage of the sum-frequency resonance is that it can be combined with a low-pass noise filter that mitigates noise at the qubit Larmor frequency, but enables control with lower frequencies, as demonstrated in Ref. \onlinecite{schirk2024protected}.
 
We note that all of our formulas for the Rabi frequencies and the Bloch-Siegert shifts, derived above for the sum-frequency resonance, can be applied to the difference-frequency resonance as well, by replacing  $\omega_2$ with $-\omega_2$. A special case of the difference-frequency resonance is the Raman resonance \cite{hays2021coherent,john2024bichromatic} when both drive frequencies are tuned slightly off-resonance from an excited state of the system. Although the Rabi frequency can be boosted in this case, during the calculations we assumed that the driving frequencies are much smaller than the energy difference between the qubit states and the excited states, making our analytical results invalid for Raman transitions.

\section{Conclusion}
\label{summary}

In this paper, we investigated the $g$-matrix formalism of semiconductor spin qubits, which was known to be valid in the case of monochromatic driving \cite{venitucci2018electrical,crippa2018electrical} with a single gate, but the question of whether it could capture more complicated dynamics was unknown. Using a general model of a spin qubit we have shown that in the case of monochromatic driving with two gates and bichromatic driving with a single gate, the $g$-TF yields the correct dynamics, and we expressed the Rabi frequency and the Bloch-Siegert shift using the $g$-tensor and its derivatives. This means that during the derivation of the effective qubit Hamiltonian, the time-dependence of the driving electric field does not have to be considered, it is enough to replace a static electric field with a modulated one in the effective Hamiltonian. However, this is not true when two different gates are used for bichromatic driving, in this case extra, all-electric terms have to be taken into account, which are not captured by the $g$-tensor.  
We demonstrated the general findings using a concrete model by comparing results obtained from $g$-TF, analytical calculations, or numerical simulations, assuming an electron (hole) trapped in a 2D harmonic potential, with an in-plane magnetic field and Rashba spin-orbit interaction.

The $g$-matrix formalism is a convenient and commonly used method to describe spin-qubit dynamics simply, therefore, it is important to understand its limitations. The question is relevant not only in semiconductor spin-based quantum computing, it is important whenever we study a driven quantum system and would like to describe it using an effective, low-dimensional Hamiltonian. An example of such a system is the fluxonium qubit, where non-trivial subharmonic driving \cite{schirk2024protected} or simultaneous charge and flux pulsing \cite{rower2024suppressing, sank2025balanced} can lead to reduced operation errors. 
\begin{acknowledgments}

We acknowledge fruitful discussions with B. Kolok. This research was supported
by the Ministry of Culture and Innovation, and the National Research, Development and Innovation Office within the Quantum Information National Laboratory of Hungary (Grant No. 2022-2.1.1-NL-2022-00004), by the J\'{a}nos Bolyai Research Scholarship of the Hungarian Academy of Science, and by the NKFIH through the OTKA Grants FK 134437. This research was supported by the European Union within the Horizon Europe research and innovation programme via the IGNITE, ONCHIPS, and QLSI2 projects, and by the HUN-REN Hungarian Research Network through the Supported Research Groups Programme, HUN-REN-BME-BCE Quantum Technology Research Group (TKCS-2024/34). Z.Gy. was supported by the EKÖP-24 University Excellence Scholarship Program of the Ministry for Culture 
and Innovation from the source of the National Research, Development and Innovation fund.
\end{acknowledgments}

\appendix

\section{Time-dependent Schrieffer-Wolff transformation}
\label{appendix_time_SW}
Schrieffer-Wolff transformation 
\cite{schrieffer1966relation,bravyi2011schrieffer} is often used to derive an effective low-dimensional Hamiltonian that describes the investigated quantum system. If the original Hamiltonian is time-dependent, then the time-dependent Schrieffer-Wolff transformation (TDSW) has to be applied instead \cite{romhanyi2015subharmonic}. Here we quickly review the TDSW and give the results up to third order, more details can be found in Ref. \onlinecite{romhanyi2015subharmonic}.

According to TDSW, we apply on the original (in our case infinite-dimensional) Hamiltonian a time-dependent unitary transformation $U(t)$, which preserves the form of the time-dependent Schrödinger-equation: 
\begin{equation}
    \Tilde{H}(t)=UH(t)U^\dag +i \hbar \dot{U} U^\dag, 
\end{equation}
where $\dot{U}$ denotes the time-derivative of $U$. The transformed Hamiltonian $\Tilde{H}$ becomes block-diagonal up to a given order in the perturbation. One block consists of the states of the effective Hamiltonian $H_\mathrm{eff}$ (i.e. relevant states), and the other block contains all the irrelevant states. In our example, the relevant states are the qubit states, and the irrelevant states are all the excited orbital states. The operator $U$ is written in an exponential form and the exponent is written as a perturbative series: 

\begin{equation}
    U=\mathrm{e}^{-S}, \hspace{5mm} S=S_1+S_2+\hdots, 
\end{equation}
where $S$ is a time-dependent anti-Hermitian operator and $S_i$ denotes the $i$th perturbative correction to $S$. 

The original Hamiltonian $H(t)$ is divided into two parts, the unperturbed $H^{(0)}$ and the perturbation $H_P$, $H=H^{(0)}+H_P$ and the spectrum of $H^{(0)}$ is assumed to be known. Furthermore, it is convenient to write the perturbation $H_P$ as the sum of a block-diagonal matrix $H_1$ and a block-off diagonal $H_2$, $H_P=H_1+H_2$.  

Upon periodic driving of the system, e.g. monochromatic driving, the time-derivative $\dot{S}_i$ of $S_i$ is in the same order of magnitude as $\omega S_i$, where $\omega$ is the driving frequency. We assume that $\hbar\omega$ has the same order of magnitude as the perturbation, therefore $\dot{S}_i$ is in the order of $S_{i+1}$. This statement will also be true in the case of bichromatic driving with frequencies $\omega_1$ and $\omega_2$, if both driving frequencies have the same order of magnitudes as the perturbation. 

The effective Hamiltonian $H_\mathrm{eff}$ can be written as a perturbative series, $H_\mathrm{eff}=H^{(0)}+H^{(1)}+H^{(2)}+\hdots$. If we prescribe that all terms in the series are block-diagonal, we obtain equations for the $S_i$ operators. The results up to $S_2$: 
\begin{subequations}
\begin{align}
    [H^{(0)},S_1] &= -H_2, \label{eq:S1}\\
    [H^{(0)},S_2] &= -[H_1,S_1]+i\hbar \dot{S}_1 ,\label{eq:S2}\\
    &\vdots \notag
\end{align}
\end{subequations}
The first equation can be solved algebraically for $S_1$, then the solution can be inserted in the second equation to obtain $S_2$. Using these, the terms in the effective Hamiltonian can be determined: 
\begin{subequations}
\begin{align}
    H^{(1)} &= H_1,\\
    H^{(2)} &= \frac{1}{2}[H_2,S_1] ,\label{eq:Heff2}\\
    H^{(3)} &= \frac{1}{2}[H_2,S_2] ,\label{eq:Heff3}\\
    &\vdots \notag
\end{align}
\end{subequations}
We can see that $S_2$ already contains time derivatives, therefore starting from the third-order correction $H^{(3)}$ of the effective Hamiltonian, the time derivatives appear and cannot be neglected. 

\section{General model and proof}
\label{app_general_model} In Section \ref{Rashba_model} we have seen that in the case of a bichromatically driven circular quantum dot with Rashba spin-orbit interaction, the $g$-tensor formalism works if the driving is realized via a single gate. If we apply bichromatic driving using two gates, then $g$-TF fails. Here we show that this is generally true if some feasible conditions are fulfilled. 

Let us consider the general infinite-dimensional qubit Hamiltonian from Eq.~\eqref{eq:H}.  We consider the kinetic, confinement, and spin-orbit terms the unperturbed Hamiltonian $H^{(0)}=H_\textrm{kin}+H_\textrm{conf}(V_g^{(0)})+H_\textrm{SO}$, while the magnetic and electric terms will be treated as perturbations $H_P=H_B+H_\textrm{conf}(V_g^{(0)}+\delta V(t))-H_\mathrm{conf}(V_g^{(0)})$, so that $H(t)=H^{(0)}+H_P$. The driving potential can be expanded around the static gate voltage $V_g^{(0)}$, first assuming a single driving gate is used: 
\begin{equation}\label{eq:conf_expansion}
\begin{aligned}
    &H_\mathrm{conf}(V_g^{(0)}+\delta V(t))-H_\mathrm{conf}(V_g^{(0)})=\\& 
    H'_\mathrm{conf}(V_g^{(0)})\delta V(t)+\frac{1}{2}H''_\mathrm{conf}(V_g^{(0)})\delta V(t)^2+\hdots 
\end{aligned}
\end{equation}
For the sake of simplicity, we introduce the following notation for the first-order term from Eq.~\eqref{eq:conf_expansion}: 
\begin{equation}\label{eq:HdeltaV}
    H_{\delta V}=H'_\mathrm{conf}(V_g^{(0)})\delta V(t)=Q_{\delta V}\delta V(t), 
\end{equation}
where $Q_{\delta V}=H'_\mathrm{conf}(V_g^{(0)})$. We also introduce $H_{\delta V^2}$ for the second-order term: 
\begin{equation}
   H_{\delta V^2}=\frac{1}{2}H''_\mathrm{conf}(V_g^{(0)})\delta V(t)^2=\frac{1}{2}Q_{\delta V^2}\delta V^2,
\end{equation}
where we introduced the notation $Q_{\delta V^2}=H''_\mathrm{conf}(V_g^{(0)})$.

To obtain an effective Hamiltonian that can describe the bichromatic dynamics, we have to apply TDSW up to the third order. To capture all leading-order contributions, terms up to the second order from the expansion of the confinement potential Eq.~\eqref{eq:conf_expansion} must be included in the perturbative calculation. This means that the perturbation is $H_P=H_B+H_{\delta V}+H_{\delta V^2}$, which is different from the standard TDSW described in App. \ref{appendix_time_SW}. $H_{\delta V^2}$ is a correction to $H_{\delta V}$ and $H_B$, hence $H_P$ contains terms of two different orders of magnitude (we assume $H_{\delta V}$ and $H_B$ are comparable). Nevertheless, TDSW can be applied, but a complication appears regarding the order of different terms. In the case of bichromatic driving, terms involving $H_{\delta V^2}$ from the second-order TDSW can have the same order of magnitude as terms containing $H_{\delta V}$, obtained from the third-order TDSW. 

 We work on the eigenbasis of the unperturbed Hamiltonian $H^{(0)}$, see Eq.~\eqref{eq:basis}. 
  We introduce the following notation for the energy difference between the excited states and the lowest Kramers pair:
\begin{equation}
    E_{k}-E_{0}=\Delta _k.
\end{equation}

On the eigenbasis of $H^{(0)}$ we can identify $H_1$ and $H_2$ as the block-diagonal and off-diagonal perturbations, which come from either the driving potential or the magnetic field. The first-order driving Hamiltonian $H_{\delta V}$ is invariant under time-reversal symmetry (TRS), this means that the following relations will be true for the matrix elements of $H_{\delta V}$: 
\begin{equation}\label{eq:time-reversal}
    H_{\delta V,k\uparrow l\downarrow}=-H_{\delta V,l\uparrow k\downarrow}, \hspace{5mm}
    H_{\delta V,k\uparrow l\uparrow}=H_{\delta V,l\downarrow k\downarrow},
\end{equation}
where we used the $\bra{\Psi_{k\uparrow}}H_{\delta V}\ket{\Psi_{l\downarrow}}=H_{\delta V,k\uparrow l \downarrow}$ notation for the matrix elements. Note that Eq.~\eqref{eq:time-reversal} is true for any term of the expansion from Eq.~\eqref{eq:conf_expansion}.

We aim to derive a two-dimensional effective Hamiltonian using TDSW. The consequence of TRS is that there will be no first-order coupling between the states of the effective Hamiltonian ($\ket{\Psi_{0\uparrow}}$ and $\ket{\Psi_{0\downarrow}}$), therefore at least a second-order transformation needs to be applied. The second-order contribution to the effective Hamiltonian: 
    \begin{equation}\label{eq:TDSWsecond}
    H^{(2)}_{\alpha \beta}=-\sum\displaylimits_{l=1}^{\infty} \frac{H_{P,0\alpha l\uparrow}H_{P,l\uparrow 0\beta}+H_{P,0\alpha l \downarrow} H_{P,l\downarrow 0 \beta }}{\Delta_l}.
\end{equation}

\subsection{Monochromatic driving}
\label{appendix_mono}
First, we investigate the case of monochromatic driving described in Ref. \onlinecite{crippa2018electrical}. To capture the leading-order dynamics, only the first-order expansion term $H_{\delta V}$ (Eq.~\eqref{eq:HdeltaV}) of the confinement potential has to be taken into account from Eq.~\eqref{eq:conf_expansion}. Monochromatic driving means that only a single $H_{\delta V}$ term will contribute to the dynamics in the perturbative description. However, an electric driving field alone cannot induce transitions, the external magnetic field is required, and the dynamics will be in second order. To investigate the Rabi frequency, we have to look at the off-diagonal matrix element of the effective Hamiltonian $H_\mathrm{eff}$, in leading order the second-order correction $H^{(2)}$. Due to TRS, the second-order terms containing two driving potentials, here denoted $H^{(2)}_{\delta V\delta V}$, will cancel in the off-diagonal matrix elements: 
\begin{widetext}
    \begin{equation}\label{eq:TDSW2EE}
    H^{(2)}_{\delta V\delta V,\uparrow \downarrow}=-\sum\displaylimits_{l=1}^{\infty} \frac{H_{\delta V,0\uparrow l\uparrow}H_{\delta V,l\uparrow 0\downarrow}+H_{\delta V,0\uparrow l \downarrow} H_{\delta V,l\downarrow 0 \downarrow }}{\Delta_l}=-\sum\displaylimits_{l=1}^{\infty} \frac{H_{\delta V,l\uparrow 0\downarrow}(H_{\delta V,0\uparrow l\uparrow}-H_{\delta V,l\downarrow 0 \downarrow})}{\Delta_l}=0,
\end{equation}
\end{widetext}
where we used the identities from Eq.~\eqref{eq:time-reversal}. Terms containing two magnetic fields will be irrelevant to the dynamics because those are time-independent, therefore the Rabi frequency is linear in both magnetic and electric fields. We have to look at the diagonal matrix elements, whether two electric driving fields can give different diagonal matrix elements or not. The first diagonal matrix element is: 
\begin{equation}\label{eq:diag1}
    H^{(2)}_{\delta V\delta V,\uparrow \uparrow}=-\sum\displaylimits_{l=1}^{\infty} \frac{H_{\delta V,0\uparrow l\uparrow}H_{\delta V,l\uparrow 0\uparrow}+H_{\delta V,0\uparrow l \downarrow} H_{\delta V,l\downarrow 0 \uparrow }}{\Delta_l}, 
\end{equation}
and the second is: 
\begin{equation}\label{eq:diag2}
    H^{(2)}_{\delta V\delta V,\downarrow \downarrow}=-\sum\displaylimits_{l=1}^{\infty} \frac{H_{\delta V,0\downarrow l\uparrow}H_{\delta V,l\uparrow 0\downarrow}+H_{\delta V,0\downarrow l \downarrow} H_{\delta V,l\downarrow 0 \downarrow }}{\Delta_l}. 
\end{equation}
Based on TRS, we can see that the first (second) term of Eq.~\eqref{eq:diag1} is equal to the second (first) term of Eq.~\eqref{eq:diag2}, therefore the two diagonal corrections are equal. The remaining terms contributing to the monochromatic process:
\begin{equation}
    H^{(2)}_{\alpha\beta}=-\sum\displaylimits_{l=1}^\infty \sum\displaylimits_{\gamma=\{\downarrow,\uparrow\}}\frac{H_{\delta V,0\alpha l\gamma}H_{B,l\gamma 0\beta}+H_{B,0\alpha l\gamma}H_{\delta V,l\gamma 0\beta}}{\Delta_l}.
\end{equation}

This way it is clear that the dynamics is linear in the magnetic field. The second-order TDSW does not contain time derivatives, which means that the static effective Hamiltonian can be derived using a time-independent Schrieffer-Wolff transformation and a static gate voltage. This effective Hamiltonian can be captured by a static $g$-tensor (being linear in the magnetic field) and if we replace the static gate voltage $V_g^{(0)}$ with a modulated $V_g^{(0)}+V_\textrm{ac}\cos{\omega t}$ field, the dynamics is obtained correctly. Hence, the monochromatic $g$-TF is indeed correct if we consider only the magnetic field and driving potential as perturbations. If the spin-orbit interaction was treated perturbatively together with the driving electric field and the magnetic field, the time-independent Schrieffer-Wolff transformation could not capture the dynamics. 

We remark that here we assumed that monochromatic driving is realized using a single driving gate. This is not a requirement, when two driving gates are used also a second-order time-independent Schrieffer-Wolff transformation describes the dynamics, therefore $g$-TF is valid for monochromatic driving using two gates. 

\subsection{Bichromatic driving with one gate}\label{App_general_oneop}
Here we show that the $g$-TF can correctly describe bichromatic driving using a single gate. Furthermore, we show the calculation required to express the Rabi frequency Eq.~\eqref{eq:gTMR_Rabi_bi} and the Bloch-Siegert shift Eq.~\eqref{eq:gTMR_BS} using the $g$-tensor and its derivatives. 

\subsubsection{Derivation of the time-dependent effective Hamiltonian}\label{App:oneop_Heff}
The effective Hamiltonian describing bichromatic driving using a single gate is of third order: 
\begin{equation}
    H_\mathrm{eff}=H^{(0)}+H^{(1)}+H^{(2)}+H^{(3)}.
\end{equation}
The matrix elements of the 0th order term are $H_{\alpha\beta}^{(0)}=E_0\delta_{\alpha\beta}$. The matrix elements of the first-order term are $H_{\alpha\beta}^{(1)}=H_{B,0\alpha0\beta}+H_{\delta V,0\alpha 0\beta}+H_{\delta V^2,0\alpha 0\beta}$. The driving terms $H_{\delta V}$ and $H_{\delta V^2}$ cannot influence the dynamics due to TRS, therefore only the magnetic term remains in $H^{(1)}$. 

We have seen that second-order dynamics involving only two electric driving fields is not possible due to TRS. A bichromatic process requires two driving electric fields, which in leading order can come from second- and third-order terms from TDSW, $H^{(2)}$ and $H^{(3)}$. The second-order term $H^{(2)}$ comes from Eq.~\eqref{eq:TDSWsecond}, with $H_P$ perturbation $H_P=H_B+H_{\delta V}+H_{\delta V^2}$: 
\begin{equation}\label{eq:bichro_second}
\begin{aligned}
    H_{\alpha\beta}^{(2)}=&-\sum\limits_{l,\eta}\frac{H_{\delta V,0\alpha l\eta}H_{B,l\eta 0\beta}+H_{B,0\alpha l\eta}H_{\delta V,l\eta 0\beta}}{\Delta_l}\\
    &-\sum\limits_{l,\eta}\frac{H_{\delta V^2,0\alpha l\eta}H_{B,l\eta 0\beta}+H_{B,0\alpha l\eta}H_{\delta V^2,l\eta 0\beta}}{\Delta_l},
\end{aligned}
\end{equation}
where the summation over $l$ is understood from 1 to $\infty$, and $\eta=\{\uparrow,\downarrow\}$. The second sum from Eq.~\eqref{eq:bichro_second} containing $H_{\delta V^2}$ contributes to the bichromatic process via the term proportional to $\cos{\omega_1 t}\cos{\omega_2 t}$, as $H_{\delta V^2}\propto \delta V^2$ (see Eq.~\eqref{eq:modulation_single}). $H^{(2)}$ is linear in the magnetic field and can be obtained using a time-independent Schrieffer-Wolff transformation, therefore it is incorporated by $g$-TF.

Another leading-order contribution to the bichromatic process appears from third-order TDSW, considering only the linear term $H_{\delta V}$ of the confinement potential, $H_P=H_B+H_{\delta V}$. The third-order correction $H^{(3)}$ to the effective Hamiltonian from TDSW: 
\begin{widetext}
    \begin{equation}\label{eq:TDSW3}
    \begin{aligned}
        H_{\alpha\beta}^{(3)}=& \sum\limits_{l,\eta,m,\gamma}\frac{H_{P,0\alpha l\eta}H_{P,l\eta m\gamma}H_{P,m \gamma 0\beta}}{\Delta_l \Delta_{m}}-\frac{1}{2}\sum\limits_{l,\eta,\gamma}\frac{H_{P,0\alpha l\eta}H_{P,l\eta 0\gamma}H_{P,0\gamma 0\beta}}{\Delta_l^2} +\\
        &-\frac{1}{2}\sum\limits_{l,\eta,\gamma}\frac{H_{P,0\alpha 0\gamma}H_{P,0\gamma l\eta}H_{P,l\eta 0\beta}}{\Delta_l^2}+\frac{i\hbar}{2}\sum\limits_{l,\eta}\frac{\Dot{H}_{P,0\alpha l\eta}H_{P,l\eta 0\beta}-H_{P,0\alpha l\eta}\Dot{H}_{P,l\eta 0\beta}}{\Delta_l^2}, 
    \end{aligned}
\end{equation}
\end{widetext}
where the summations over $l$ and $m$ are understood from 1 to $\infty$, and the summations over $\eta$ and $\gamma$ from $\uparrow$ to $\downarrow$. 

If the last sum of Eq.~\eqref{eq:TDSW3} containing the time derivatives vanished, then the remaining terms relevant to the dynamics could be captured by a time-independent Schrieffer-Wolff transformation. In the remaining, first three terms of Eq.~\eqref{eq:TDSW3} exactly two driving terms $H_{\delta V}$ have to appear to contribute to the bichromatic process. This means that the third $H_P$ term must be magnetic, this way the effective Hamiltonian is linear in the magnetic field and does not contain time derivatives, $g$-TF will be valid. We only keep those terms of the first three sums of Eq.~\eqref{eq:TDSW3} that are linear in the magnetic field, as these are relevant for bichromatic driving.

When the driving is realized using a single driving gate, the first-order driving term can be written as: 
\begin{equation}
    H_{\delta V}=(V_\mathrm{ac,1}\sin{\omega_1 t}+V_\mathrm{ac,2}\sin{\omega_2 t})Q_{\delta V}, 
\end{equation}
This means that the time derivative of $H_{\delta V}$ will be proportional to the same $Q_{\delta V}$, so: 
\begin{equation}\label{eq:derivative_proportional}
    \dot{H}_{\delta V} \propto H_{\delta V}. 
\end{equation}
therefore the last sum of Eq.~\eqref{eq:TDSW3} containing time derivatives will be proportional to the second-order terms in Eq.~\eqref{eq:TDSW2EE}, which we have seen have zero off-diagonal and equal diagonal matrix elements, therefore irrelevant. Hence, the time derivatives vanish, $g$-TF is also valid in the bichromatic case when the driving is achieved using a single gate. 

Note that here we used the eigenbasis of the unperturbed Hamiltonian $H^{(0)}$, which is not the eigenbasis of the qubit. A strong intrinsic spin-orbit interaction together with the external magnetic field can result in offdiagonal perturbations (the $g$-tensor is not diagonal). However, the arguments that we have used to prove that the time-derivative terms are irrelevant are still valid, because the diagonal perturbation coming from time derivatives is diagonal on the qubit's eigenbasis as well. 

We remark that we assumed that the second sum from Eq.~\eqref{eq:bichro_second} and the third-order terms from Eq.~\eqref{eq:TDSW3} (with $H_P=H_B+H_{\delta V}$) have the same order of magnitude. This is because $H_{\delta V}$ is the first-order, while $H_{\delta V^2}$ is the second-order term from the expansion of the modulated confinement potential from Eq.~\eqref{eq:conf_expansion}. We proved that $g$-TF contains both, therefore bichromatic driving using a single driving gate can be described using $g$-TF. 

We have proved that the correct effective Hamiltonian can be obtained using time-independent Schrieffer-Wolff transformation. Additionally, we have to see that the effective Hamiltonian can be expressed using the $g$-tensor and its derivatives, as in Eq.~\eqref{eq:effective}. The effective Hamiltonian has the following form: 
\begin{equation}\label{eq:Heff_bichro_oneop_form}
    H_\mathrm{eff}=H_\mathrm{eff}^{(0)}+Q_\mathrm{eff}^{(\delta V)}\delta V+\frac{1}{2}Q_\mathrm{eff}^{(\delta V^2)}\delta V^2, 
\end{equation}
where $Q_\mathrm{eff}^{(\delta V)}$ and $Q_\mathrm{eff}^{(\delta V^2)}$ are self-adjoint operators. In this case, we can Taylor-expand the effective Hamiltonian (the $g$-tensor) from Eq.~\eqref{eq:H_efft} up to the second order, and obtain the correct effective Hamiltonian Eq.~\eqref{eq:Heff_bichro_oneop_form}. The matrix elements of $H_\mathrm{eff}^{(0)}$ are: 
\begin{equation}
    H_\mathrm{eff,\alpha\beta}^{(0)}=E_0\delta_{\alpha\beta}+H_{B,0\alpha 0\beta}. 
\end{equation}
The matrix elements of $Q_\mathrm{eff}^{(\delta V)}$: 
\begin{equation}\label{eq:QeffdeltaV}
    Q_\mathrm{eff,\alpha\beta}^{(\delta V)}=-\sum_{l,\eta}\frac{Q_{\delta V,0\alpha l\eta}H_{B,l\eta 0\beta}+H_{B,0\alpha l\eta}Q_{\delta V,l\eta 0\beta}}{\Delta_l}.
\end{equation}
The matrix elements of $Q_\mathrm{eff}^{(\delta V^2)}$: 
\begin{widetext}
\begin{equation}\label{eq:QeffdeltaV2}
\begin{aligned}
    Q_\mathrm{eff,\alpha\beta}^{(\delta V^2)}&=-\sum_{l,\eta}\frac{Q_{\delta V^2,0\alpha l\eta}H_{B,l\eta 0\beta}+H_{B,0\alpha l\eta}Q_{\delta V^2,l\eta 0\beta}}{\Delta_l}+\sum_{l,\eta,m,\gamma}\frac{2}{\Delta_{l}\Delta_m}\left(H_{B,0\alpha l\eta}Q_{\delta V,l\eta m\gamma}Q_{\delta V,m\gamma 0\beta}+\right.\\&\left.+ Q_{\delta V,0\alpha l\eta}H_{B,l\eta m \gamma}Q_{\delta V,m\gamma 0\beta}+Q_{\delta V,0\alpha l\eta}Q_{\delta V,l\eta m\gamma}H_{B,m\gamma 0\beta}\right)
    -\sum_{l,\eta,\gamma}\frac{1}{\Delta_l^2}\left(H_{B,0\alpha l\eta}Q_{\delta V,l\eta 0\gamma}Q_{\delta V,0\gamma 0\beta}+\right.\\&\left.+Q_{\delta V,0\alpha l\eta}H_{B,l\eta 0\gamma}Q_{\delta V,0\gamma 0\beta}+Q_{\delta V,0\alpha l\eta}Q_{\delta V,l\eta 0\gamma}H_{B,0\gamma 0\beta} \right)-\sum_{l,\eta,\gamma}\frac{1}{\Delta_l^2}\left(H_{B,0\alpha 0\gamma}Q_{\delta V,0\gamma l\eta}Q_{\delta V,l\eta 0\beta}+\right.\\&\left.+Q_{\delta V,0\alpha 0\gamma}H_{B,0\gamma l\eta}Q_{\delta V,l\eta 0\beta}+Q_{\delta V,0\alpha 0\gamma}Q_{\delta V,0\gamma l\eta}H_{B,l\eta 0\beta}\right).
\end{aligned}
\end{equation}
\end{widetext}
The relationship between the effective Hamiltonian and the $g$-tensor: 
\begin{equation}\label{eq:Heff_gtensor}
    H_\mathrm{eff}=\frac{1}{2}\mu_B B \sigma_i g_{ij} b_j, 
\end{equation}
where $b$ is the magnetic unit vector and summation over $i$ and $j$ is understood from 1 to 3. We can express the $g$-tensor if we multiply Eq.~\eqref{eq:Heff_gtensor} from the left with $\sigma_k$ and calculate the trace: 
\begin{equation}
    \mathrm{Tr}(\sigma_k H_\mathrm{eff})=\frac{1}{2}\mu_B B g_{ij}b_j \mathrm{Tr}(\delta_{ik}I+i\epsilon_{kil}\sigma_l)=\mu_B B g_{kj}b_j,
\end{equation}
where we used the $\sigma_k \sigma_i=\delta_{ik}I+i \epsilon_{kil}\sigma_l$ identity of Pauli-matrices, where $I$ is the $2\times 2$ identity matrix, $\delta$ is the Kronecker-delta and $\epsilon$ is the Levi-Civita tensor. The $g$-tensor acting on vector $b$ this way becomes: 
\begin{equation}\label{eq:gtensor_Heff}
    g_{kj}b_j=\frac{1}{\mu_B B} \mathrm{Tr}(\sigma_k H_\mathrm{eff}). 
\end{equation}
Using $Q_\mathrm{eff}^{(\delta V)}$ from Eq.~\eqref{eq:QeffdeltaV}, $Q_\mathrm{eff}^{(\delta V^2)}$ from Eq.~\eqref{eq:QeffdeltaV2} and $H_\mathrm{eff}$ from Eq.~\eqref{eq:Heff_bichro_oneop_form}, the $g$-tensor's first and second derivatives can be evaluated at $V_{g}^{(0)}$: 
\begin{subequations}
\begin{align}
    g'_{kj}(V_g^{(0)})b_j&=\frac{1}{\mu_BB}\sum_{\alpha,\beta}\sigma_{k,\beta\alpha}Q_\mathrm{eff,\alpha \beta}^{(\delta V)}\\
    g''_{kj}(V_g^{(0)})b_j&=\frac{1}{\mu_BB}\sum_{\alpha,\beta}\sigma_{k,\beta\alpha}Q_\mathrm{eff,\alpha \beta}^{(\delta V^2)}, 
\end{align}
\end{subequations}
where summation over index $j$ is understood from 1 to 3. 

\subsubsection{Derivation of the Rabi frequency and the Bloch-Siegert shift}

To derive Eq.~\eqref{eq:gTMR_Rabi_bi} and Eq.~\eqref{eq:gTMR_BS} describing the Rabi frequency and the Bloch-Siegert shift, we start with the Hamiltonian from Eq.~\eqref{eq:effective}. The bichromatic modulation $\delta V(t)$ corresponding to driving with a single gate can be found in Eq.~\eqref{eq:modulation_single}. The Hamiltonian can be rewritten with the notations introduced in Section \ref{gTMR_bichro}:
\begin{equation}\label{eq:App_gTMR_Ham}
H=\hbar\bm{\Omega}\cdot\bm{\sigma}/2+\hbar\bm{\Omega}'\cdot\bm{\sigma}\delta V+\hbar\bm{\Omega}''\cdot\bm{\sigma}\delta V^2. 
\end{equation}
The first term of Eq.~\eqref{eq:App_gTMR_Ham}, $\hbar\bm{\Omega}\cdot\bm{\sigma}/2$ describes the static Hamiltonian of the qubit, which we diagonalize. We parametrize vector $\bm{\Omega}$ using spherical polar angles $\theta$ and $\varphi$:
\begin{equation}
    \bm{\Omega}=\Omega \begin{pmatrix}
    \sin{\theta}\cos{\varphi} \\ 
    \sin{\theta}\sin{\varphi} \\ 
    \cos{\theta}  
    
   \end{pmatrix} =\Omega \Hat{\bm{r}},
\end{equation}
where $\Hat{\bm{r}}$ is a unit vector.
We can write vectors $\bm{\Omega}'$ and $\bm{\Omega}''$ as a sum of components parallel and perpendicular to $\bm{\Omega}$:
\begin{equation}
    \bm{\Omega}'=\bm{\Omega}'_\parallel+\bm{\Omega}'_\perp, \hspace{5mm} \bm{\Omega}''=\bm{\Omega}''_\parallel+\bm{\Omega}''_\perp. 
\end{equation}
The parallel components can be expressed using unit vector $\Hat{\bm{r}}$: 
\begin{equation}
    \bm{\Omega}'_\parallel=\Omega'_\parallel \Hat{\bm{r}}, \hspace{5mm} \bm{\Omega}''_\parallel=\Omega''_\parallel \Hat{\bm{r}},
\end{equation}
where $\Omega'_\parallel$ and $\Omega''_\parallel$ are the absolute values of vectors $\bm{\Omega}'_\parallel$ and $\bm{\Omega}''_\parallel$. The perpendicular components can be written using spherical unit vectors $\Hat{\bm{\theta}}$ and $\Hat{\bm{\varphi}}$:
\begin{equation}
\begin{aligned}    &\bm{\Omega}'_\perp=\Omega'_\perp(\cos{\chi_1}\Hat{\bm{\theta}}+\sin{\chi_1}\Hat{\bm{\varphi}}), \\& \bm{\Omega}''_\perp=\Omega''_\perp(\cos{\chi_2}\Hat{\bm{\theta}}+\sin{\chi_2}\Hat{\bm{\varphi}}),
\end{aligned}
\end{equation}
where angles $\chi_1$ and $\chi_2$ were introduced and the unit vectors are: \begin{equation}
    \Hat{\bm{\theta}}=\begin{pmatrix}
        \cos{\theta}\cos{\varphi} \\ 
        \cos{\theta}\sin{\varphi}\\
        -\sin{\theta}\end{pmatrix}, \hspace{5mm}
        \Hat{\bm{\varphi}}=\begin{pmatrix}
            -\sin{\varphi} \\ 
            \cos{\varphi} \\
            0
    \end{pmatrix}.
\end{equation} 

If we transform our basis to the eigenbasis of the static qubit Hamiltonian $\hbar\bm{\Omega}\cdot \bm{\sigma}/2$, we arrive at the following time-dependent Hamiltonian: 

\begin{widetext}
\begin{equation}
    H=\hbar\begin{pmatrix}
        \Omega/2+\Omega'_\parallel \delta V+\Omega''_\parallel \delta V^2 & \Omega'_\perp \mathrm{e}^{-i(\chi_1+\varphi)}\delta V+\Omega''_\perp \mathrm{e}^{-i(\chi_2+\varphi)}\delta V^2 \\[1mm] 
        \Omega'_\perp \mathrm{e}^{i(\chi_1+\varphi)}\delta V+\Omega''_\perp \mathrm{e}^{i(\chi_2+\varphi)}\delta V^2 & -\Omega/2-\Omega'_\parallel \delta V-\Omega''_\parallel \delta V^2
    \end{pmatrix}.
\end{equation}    
\end{widetext}

We use two-mode Floquet theory \cite{ho1983semiclassical} to derive the Rabi frequency and Bloch-Siegert shift. The elements of the Floquet matrix $H_F$: 

\begin{widetext}

\begin{equation}
    \begin{aligned}
        &\bra{\alpha n_1 n_2}H_F\ket{\beta k_1 k_2}=\frac{\hbar\Omega}{2}\sigma_{z,\alpha\beta}\delta_{n_1k_1}\delta_{n_2k_2}+ (n_1 \hbar \omega_1+n_2 \hbar\omega_2)\delta_{\alpha \beta}\delta_{n_1k_1}\delta_{n_2k_2}+\left(\frac{V_\mathrm{ac,1}^2}{2}+\frac{V_\mathrm{ac,2}^2}{2} \right)\delta_{n_1k_1}\delta_{n_2k_2}V_{2,\alpha\beta}+\\& 
        +\frac{V_\mathrm{ac,1}V_\mathrm{ac,2}}{2}(\delta_{n_1-k_1,1}\delta_{n_2-k_2,1}+\delta_{n_1-k_1,-1}\delta_{n_2-k_2,-1})V_{2,\alpha\beta} +\frac{V_\mathrm{ac,1}V_\mathrm{ac,2}}{2}(\delta_{n_1-k_1,1}\delta_{n_2-k_2,-1}+\delta_{n_1-k_1,-1}\delta_{n_2-k_2,1})V_{2,\alpha\beta}+\\&
        +\frac{V_\mathrm{ac,1}}{2}(\delta_{n_1-k_1,1}+\delta_{n_1-k_1,-1})\delta_{n_2k_2}V_{1,\alpha\beta}+\frac{V_\mathrm{ac,2}}{2}(\delta_{n_2-k_2,1}+\delta_{n_2-k_2,-1})\delta_{n_1k_1}V_{1,\alpha\beta}+ \\&+\frac{V_\mathrm{ac,1}^2}{4}(\delta_{n_1-k_1,2}+\delta_{n_1-k_1,-2})\delta_{n_2k_2}V_{2,\alpha\beta} + 
        \frac{V_\mathrm{ac,2}^2}{4}(\delta_{n_2-k_2,2}+\delta_{n_2-k_2,-2})\delta_{n_1k_1}V_{2,\alpha\beta}, 
    \end{aligned}
\end{equation}

\end{widetext}
where $\alpha$ and $\beta$ are pseudospin states, $n_1$ ($k_1$) and $n_2$ ($k_2$) are photon numbers corresponding to frequency $\omega_1$ and $\omega_2$ respectively, and $V_1$ and $V_2$ are the following matrices: 
\begin{equation}
\begin{aligned}
    &V_1=\hbar\begin{pmatrix}
        \Omega'_\parallel & \Omega'_\perp \mathrm{e}^{-i(\chi_1+\varphi)} \\[1mm] 
        \Omega'_\perp \mathrm{e}^{i(\chi_1+\varphi)} & -\Omega'_\parallel 
    \end{pmatrix}, \\ 
    &V_2=\hbar\begin{pmatrix}
        \Omega''_\parallel & \Omega''_\perp \mathrm{e}^{-i(\chi_2+\varphi)} \\[1mm] 
        \Omega''_\perp \mathrm{e}^{i(\chi_2+\varphi)} & -\Omega''_\parallel
        \end{pmatrix}. 
\end{aligned}
\end{equation}

Next, we apply a second-order time-independent Schrieffer-Wolff transformation on $H_F$, we project the matrix onto states $\ket{{\uparrow} 00}$ and $\ket{{\downarrow} 11}$, the result will be the effective Floquet-matrix: 
\begin{equation}\label{eq:HFeff}
    H_\mathrm{F,eff}=\hbar\begin{pmatrix}
        \omega_\mathrm{split,1}+\omega_\mathrm{BS,1} & \pi f_\mathrm{Rabi} \\[2mm] 
        \pi f_\mathrm{Rabi}^* & \omega_\mathrm{split,2}+\omega_1+\omega_2+\omega_\mathrm{BS,2} 
    \end{pmatrix}. 
\end{equation}
Using this matrix the complex Rabi frequency $f_\mathrm{Rabi}$ can be read off, the Bloch-Siegert shift is $\omega_\mathrm{BS}=\omega_\mathrm{BS,1}-\omega_\mathrm{BS,2}$, and the splitting frequency is $\omega_\mathrm{split}=\omega_\mathrm{split,1}-\omega_\mathrm{split,2}$. To distinguish the terms contributing to the splitting frequency from those contributing to the Bloch-Siegert shift we note that the latter depend on the driving potentials $V_\mathrm{ac,1}$ and $V_\mathrm{ac,2}$. The Rabi frequency and the Bloch-Siegert shift contain higher-order contributions as well,
to obtain the leading-order results (Eq.~\eqref{eq:gTMR_Rabi_bi} and Eq.~\eqref{eq:gTMR_BS}) we need to impose the conditions $
\Omega \gg \Omega' V_\mathrm{ac,1(2)} \gg  \Omega'' V_\mathrm{ac,1(2)}^2
$ and neglect the higher-order terms. 

\subsection{Bichromatic driving with two gates}\label{App_bichro_two}
Here we prove that the $g$-TF fails to describe bichromatic driving using two different driving gates. We also express the Rabi frequency and the Bloch-Siegert shift using the $g$-tensor and the additional $\bm{\Upsilon}$ vector.

\subsubsection{Derivation of the time-dependent effective Hamiltonian}

When bichromatic driving is realized using two gates, the confinement potential can be expanded around the static gate voltages $V_{g1}^{(0)}$ and $V_{g2}^{(0)}$: 
\begin{widetext}
    \begin{equation}\label{eq:confinement_twogate}
    \begin{aligned}
        H_\mathrm{conf}(V_{g1}^{(0)}+\delta V_1(t),V_{g2}^{(0)}+\delta V_2(t))-H_\mathrm{conf}(V_{g1}^{(0)},V_{g2}^{(0)})=&\frac{\partial H_\mathrm{conf}}{\partial V_{g1}}\delta V_1(t)+\frac{\partial H_\mathrm{conf}}{\partial V_{g2}}\delta V_2(t)+\frac{1}{2}\frac{\partial^2 H_\mathrm{conf}}{\partial V_{g1}^2}\delta V_1(t)^2+ \\&+
        \frac{1}{2}\frac{\partial^2 H_\mathrm{conf}}{\partial V_{g2}^2}\delta V_2(t)^2+\frac{\partial^2 H_\mathrm{conf}}{\partial V_{g1}\partial V_{g2}}\delta V_1(t)\delta V_2(t)+\hdots
    \end{aligned}
    \end{equation}
\end{widetext}
where the partial derivatives are evaluated at $V_{g1}=V_{g1}^{(0)}$ and $V_{g2}=V_{g2}^{(0)}$. We introduce the following notation for the first-order terms from Eq.~\eqref{eq:confinement_twogate}: 
\begin{equation}\label{eq:HdeltaV_twogate}
    H_{\delta V}=\frac{\partial H_\mathrm{conf}}{\partial V_{g1}}\delta V_1(t)+\frac{\partial H_\mathrm{conf}}{\partial V_{g2}}\delta V_2(t).
\end{equation}
By introducing $Q_1$ and $Q_2$ for the partial derivatives from Eq.~\eqref{eq:HdeltaV_twogate}, $H_{\delta V}$ can be rewritten as:
\begin{equation}\label{eq:Hdeltav}
    H_{\delta V}=V_\mathrm{ac,1}\sin{(\omega_1 t)}Q_{1}+V_\mathrm{ac,2}\sin{(\omega_2 t)}Q_{2}.
\end{equation}
We introduce $H_{\delta V^2}$ for the second-order terms from Eq.~\eqref{eq:confinement_twogate}:
\begin{equation}\label{eq:HdeltaV2_twogate}
\begin{aligned}
    H_{\delta V^2}=&\frac{1}{2}\frac{\partial^2 H_\mathrm{conf}}{\partial V_{g1}^2}\delta V_1(t)^2+
        \frac{1}{2}\frac{\partial^2 H_\mathrm{conf}}{\partial V_{g2}^2}\delta V_2(t)^2+\\&+\frac{\partial^2 H_\mathrm{conf}}{\partial V_{g1}\partial V_{g2}}\delta V_1(t)\delta V_2(t).
\end{aligned}
\end{equation}

Similarly to bichromatic driving using a single gate, the leading-order dynamics has a second-order and a third-order contribution coming from TDSW, $H^{(2)}$ and $H^{(3)}$. The second-order matrix elements are the same as in Eq.~\eqref{eq:bichro_second} but using $H_{\delta V}$ from Eq. ~\eqref{eq:HdeltaV_twogate} and $H_{\delta V^2}$ from Eq.~\eqref{eq:HdeltaV2_twogate}. The third-order matrix elements can be found in Eq.~\eqref{eq:TDSW3} with the perturbation $H_P=H_B+H_{\delta V}$, $H_{\delta V}$ from Eq.~\eqref{eq:HdeltaV_twogate}.

The argument used to prove the validity of $g$-TF for bichromatic driving using a single gate is not true in this case. Namely, the statement that the time derivative of $H_{\delta V}$ is proportional to itself is not true (see Eq.~\eqref{eq:Hdeltav}), so the sum in Eq.~\eqref{eq:TDSW3} containing time derivatives is not zero. This means that the Rabi frequency will have contributions coming from time derivatives. Therefore in this case $g$-TF cannot yield the correct Rabi frequency.

To better see that the $g$-TF fails, we can express the time-dependent $g$-tensor that correctly describes the dynamics using the $g$-tensor obtained from $g$-TF and an extra correction. The $g$-tensor can be expressed using the effective Hamiltonian according to Eq.~\eqref{eq:gtensor_Heff}. The effective Hamiltonian consists of terms captured by time-independent transformation and terms coming from time derivatives. The latter cannot be obtained using the $g$-TF, therefore those have to be added as corrections:
\begin{widetext}
\begin{equation}\label{eq:g_correction}
    g(t)_{ij}b_j=g_{ij}^{g\mathrm{-TF}}(t)b_j+\frac{i\hbar}
    {2\mu_B B}\sigma_{i,\beta\alpha}\sum\limits_{l,\gamma}\frac{\Dot{H}_{\delta V,0\alpha l\gamma}H_{\delta V,l\gamma 0\beta}-H_{\delta V,0\alpha l\gamma}\Dot{H}_{\delta V,l\gamma 0\beta}}{\Delta_l^2},
\end{equation}
\end{widetext}
where summation over $\alpha$ and $\beta$ is understood, and $g^{g-\mathrm{TF}}$ is the $g$-tensor obtained from time-dependent Schrieffer-Wolff transformation. 

\subsubsection{Derivation of the Rabi frequency and the Bloch-Siegert shift}

Here we aim to calculate the Rabi frequency and the Bloch-Siegert shift. First, we calculate the Hamiltonian obtained from $g$-TF, $H_\mathrm{eff}^{g-\mathrm{TF}}(t)$ from Eq.~\eqref{eq:effective_two_punger}. We expand the $g$-tensor in Taylor series around static voltages $V_{g1}^{(0)}$ and $V_{g2}^{(0)}$ up to second order:
\begin{widetext}
\begin{equation}\label{eq:gTMR_twogate}
\begin{aligned}
    H_\mathrm{eff}^{g-\mathrm{TF}}(t)=\frac{1}{2}\mu_B \bm{\sigma}\cdot&\left[\hat{g}+\frac{\partial \hat{g}}{\partial V_{g1}}\delta V_1(t)+\frac{\partial \hat{g}}{\partial V_{g2}}\delta V_2(t)+\frac{1}{2}\frac{\partial^2\hat{g}}{\partial V_{g1}^2}\delta V_1(t)^2+\frac{1}{2}\frac{\partial^2 \hat{g}}{\partial V_{g2}^2}\delta V_2(t)^2 +\frac{\partial^2 \hat{g}}{\partial V_{g1}\partial V_{g2}}\delta V_1(t) \delta V_2(t)\right]\bm{B},
\end{aligned}
\end{equation}
\end{widetext}
where the $g$-tensor and the partial derivatives are evaluated at $V_{g1}=V_{g1}^{(0)}$ and $V_{g2}=V_{g2}^{(0)}$. The expansion of the $g$-tensor can be done similarly to the case of bichromatic driving using a single gate described in App. \ref{App:oneop_Heff}. The correct effective Hamiltonian is obtained in Eq.~\eqref{eq:gTMR_twogate} because the effective Hamiltonian itself is quadratic in modulations $\delta V_1(t)$ and $\delta V_2(t)$:
\begin{equation}
\begin{aligned}
    H_\mathrm{eff}^{g-\mathrm{TF}}=&H_\mathrm{eff}^{(0)}+Q_\mathrm{eff}^{(\delta V_1)}\delta V_1+Q_\mathrm{eff}^{(\delta V2)}\delta V_2+\frac{1}{2}Q_\mathrm{eff}^{(\delta V_1^2)}\delta V_1^2+\\
    &+\frac{1}{2}Q_\mathrm{eff}^{(\delta V_2^2)}\delta V_2^2+Q_\mathrm{eff}^{(\delta V_1,\delta V_2)}\delta V_{1}\delta V_2, 
\end{aligned}
\end{equation}
where the matrices can be read off using the terms of the effective Hamiltonian $H^{(2)}$ Eq.~\eqref{eq:bichro_second}, the first three sums of $H^{(3)}$ Eq.~\eqref{eq:TDSW3}, and the modulations $H_{\delta V}$ Eq.~\eqref{eq:HdeltaV_twogate} and $H_{\delta V^2}$ Eq.~\eqref{eq:HdeltaV2_twogate}, keeping only the terms linear in $B$. 

We introduce the following vectors:  
\begin{equation}
 \begin{aligned}
 & \hbar\bm{\Omega}=\mu_B\hat{g}\bm{B}, \hspace{3mm} \hbar\bm{\Omega}_1'=\frac{\mu_B}{2}\frac{\partial \hat{g}}{\partial V_{g1}}\bm{B}, 
 \\& \hbar\bm{\Omega}_2'=\frac{\mu_B}{2}\frac{\partial \hat{g}}{\partial V_{g2}}\bm{B}, \hspace{3mm}\hbar\bm{\Omega}_{11}''=\frac{\mu_B}{4}\frac{\partial^2 \hat{g}}{\partial V_{g1}^2}\bm{B}, \hspace{2mm} \\& \hbar\bm{\Omega}_{22}''=\frac{\mu_B}{4}\frac{\partial^2 \hat{g}}{\partial V_{g2}^2}\bm{B}, \hspace{3mm} \hbar\bm{\Omega}_{12}''=\frac{\mu_B}{2}\frac{\partial^2 \hat{g}}{\partial V_{g1}\partial V_{g2}}\bm{B}.
 \end{aligned}
\end{equation}
Here we use this notation for the partial derivatives to highlight the fact that e.g. $\bm{\Omega}$, $\bm{\Omega}_1'$ and $\bm{\Omega}_{11}''$ have different dimensions. 
We write the vectors coming from the derivatives of the $g$-tensor as a sum of components parallel and perpendicular to $\bm{\Omega}$, e.g. $\bm{\Omega}_{2}'=\bm{\Omega}_{2\parallel}'+\bm{\Omega}_{2\perp}'$. We also introduce a different angle to every vector to write the perpendicular components as a linear combination of basis vectors $\Hat{\bm{\theta}}$ and $\Hat{\bm{\varphi}}$, e.g. $\bm{\Omega}_{2\perp}'=\Omega_{2\perp}'(\cos{\chi_2}\Hat{\bm{\theta}}+\sin{\chi_2}\Hat{\bm{\varphi}})$, where $\Omega_{2\perp}'$ is the absolute value. 

Now let us turn to the other term from Eq.~\eqref{eq:effective_two_punger}, $H_\mathrm{eff}^{\mathrm{TD}}$. In Eq.~\eqref{eq:sum_formula} the $H_P$ perturbation appears, which contains magnetic and electric fields, but only the electric terms contribute to the bichromatic process, therefore instead of $H_P$ we have to write $H_{\delta V}$ from Eq.~\eqref{eq:Hdeltav}. Using the consequences of TRS (Eq.~\eqref{eq:time-reversal}) we arrive to the following expression: 
\begin{equation}\label{eq:Heff_TD}
    \begin{aligned}
        H_\mathrm{eff}^{\mathrm{TD}}(t)&=V_\mathrm{ac,1}V_\mathrm{ac,2}(\omega_1-\omega_2)\sin{[(\omega_1+\omega_2)t]}\bm{\Upsilon}\cdot \bm{\sigma}\\ 
        &-V_\mathrm{ac,1}V_\mathrm{ac,2}(\omega_1+\omega_2)\sin{[(\omega_1-\omega_2)t]}\bm{\Upsilon}
    \cdot\bm{\sigma},
    \end{aligned}
\end{equation}
where $\bm{\Upsilon}=(\Upsilon_x,\Upsilon_y,\Upsilon_z)$. 
The term proportional to $\sin{[(\omega_1+\omega_2)t]}$ will contribute to the bichromatic driving using the sum of driving frequencies, the other to the process using the difference of frequencies. We introduce the following complex number $c_Q$: 
\begin{equation}
    c_Q=\frac{i\hbar}{2}\sum\displaylimits_{l=1}^\infty \frac{Q_{1,0\uparrow l\downarrow}Q_{2,l\downarrow 0\downarrow}+Q_{1,0\uparrow l\uparrow} Q_{2,l\uparrow 0\downarrow}}{\Delta_l^2}. 
\end{equation}
The $x$ and $y$ components of vector $\bm{\Upsilon}$ can be written using the real and imaginary parts of $c_Q$: 
\begin{equation}
    \Upsilon_x=\operatorname{Re}(c_Q), \hspace{3mm} \Upsilon_y=-\operatorname{Im}(c_Q).
\end{equation}
The $z$ component can be calculated as follows: 
\begin{equation}
    \begin{aligned}
        \Upsilon_z=\frac{i\hbar}{4}\sum\displaylimits_{l=1}^\infty \frac{1}{\Delta_l^2}&\left(Q_{1,0\uparrow l\uparrow}Q_{2,l\uparrow 0\uparrow}-Q_{2,0\uparrow l\uparrow}Q_{1,l\uparrow 0\uparrow}+\right. \\+&\left. Q_{1,0\uparrow l\downarrow}Q_{2,l\downarrow 0\uparrow}-Q_{2,0\uparrow l\downarrow}Q_{1,l\downarrow 0\uparrow}\right).
    \end{aligned}
\end{equation}

We write $\bm{\Upsilon}$ as a sum of components parallel and perpendicular to $\bm{\Omega}$, $\bm{\Upsilon}=\bm{\Upsilon}_\parallel+\bm{\Upsilon}_\perp$. We further introduce $\chi_\Upsilon$ based on $\bm{\Upsilon}_\perp=\Upsilon_\perp(\cos{\chi_\Upsilon}\Hat{\bm{\theta}}+\sin{\chi_\Upsilon}\Hat{\bm{\varphi}})$, where $\Upsilon_\perp$ is the absolute value. 

To calculate the Rabi frequency and the Bloch-Siegert shift, we use the same procedure as described in App.~\ref{App_general_oneop}. We apply a basis transformation on the two terms from Eq.~\eqref{eq:gTMR_twogate} and Eq.~\eqref{eq:Heff_TD} to diagonalize the unperturbed Hamiltonian $\hbar\bm{\Omega}\cdot \bm{\sigma}/2$. Using a second-order time-independent Schrieffer-Wolff transformation, we construct the Floquet matrix and calculate an effective $2\times 2$ Floquet matrix. Finally, we read off the Rabi frequency and the Bloch-Siegert shift, see Eq.~\eqref{eq:HFeff}. The complex Rabi frequency: 
\begin{widetext}
\begin{equation}
    h f_\mathrm{Rabi}=V_\mathrm{ac,1}V_\mathrm{ac,2}e^{i\varphi}\left[\frac{e^{i\chi_2}\hbar\Omega_{1\parallel}'\Omega_{2\perp}'}{\omega_1}+\frac{e^{i\chi_1}\hbar\Omega_{2\parallel}'\Omega_{1\perp}'}{\omega_2}-\frac{e^{i\chi_{12}}\hbar\Omega_{12\perp}''}{2}- e^{i(\chi_\Upsilon+\pi/2)}(\omega_1-\omega_2) \Upsilon_\perp\right].
\end{equation}
\end{widetext}
The Bloch-Siegert shift: 
\begin{equation}
\begin{aligned}
\omega_\mathrm{BS}&=\left(\Omega_{11\parallel}''+\frac{\Omega \Omega_{1\perp}'^2}{\Omega^2-\omega_1^2}\right)V_\mathrm{ac,1}^2+\\&+\left(\Omega_{22\parallel}''+\frac{\Omega \Omega_{2\perp}'^2}{\Omega^2-\omega_2^2}\right)V_\mathrm{ac,2}^2.
\end{aligned}
\end{equation}
We can see that (for a given magnetic field direction) two extra parameters need to be included to describe the Rabi frequency, $\Upsilon_\perp$ absolute value of the perpendicular component of $\bm{\Upsilon}$ to $\bm{\Omega}$, and the $\chi_\Upsilon$ angle. The $\bm{\Upsilon}$ vector leaves the Bloch-Siegert shift unaffected. 

\section{Perturbative results}
The different Rabi frequencies of the circular quantum dot model with Rashba spin-orbit interaction were presented in Section \ref{Rashba_model}, here we provide details about the perturbative calculations necessary to arrive at those results. The Hamiltonian is written in Eq.~\eqref{eq:Rashba_Ham}, where the Zeeman term from Eq.~\eqref{eq:Zeeman} is not diagonal in spin, therefore we diagonalize it by changing the basis in the spin space. After the diagonalization the Zeeman term becomes $\Tilde{B}\sigma_z/2$, while the transformed Rashba spin-orbit interaction:   
\begin{equation}
    H_\mathrm{SO}=\alpha(p_x\Sigma_x+p_y\Sigma_y), 
\end{equation}
where $\Sigma_x$ and $\Sigma_y$ are: 
\begin{subequations}
\begin{align}
    \Sigma_x&=-\frac{1}{2}\sin{2\phi}\cdot\sigma_x+\cos^2{\phi}\cdot\sigma_y+\sin{\phi}\cdot\sigma_z \\ 
    \Sigma_y&=-\sin^2{\phi}\cdot \sigma_x+\frac{1}{2}\sin{2\phi}\cdot \sigma_y-\cos{\phi}\cdot \sigma_z. 
    \end{align}
\end{subequations}

\subsection{Bichromatic driving with one gate}\label{App_oneop}
When the qubit is driven bichromatically via a single driving gate, the electric Hamiltonian can be written as in Eq.~\eqref{eq:electric_oneop}, the resonance condition is:  
\begin{equation}\label{eq:resonance}
    \omega_1+\omega_2=\omega_\mathrm{split}+\omega_\mathrm{BS}, 
\end{equation}
where $\omega_\mathrm{split}$ is the splitting frequency of the qubit and $\omega_\mathrm{BS}$ is the Bloch-Siegert shift, which depends on the driving strengths $\Tilde{E}_\mathrm{ac_1}$ and $\Tilde{E}_\mathrm{ac_2}$. 

A fifth-order TDSW has to be applied to capture the bichromatic dynamics, the resulting effective Hamiltonian: 
\begin{equation}\label{eq:Heff_fifth}
    H_\mathrm{eff}=H^{(0)}+H^{(1)}+H^{(2)}+H^{(3)}+H^{(4)}+H^{(5)}. 
\end{equation}
$H^{(0)}$, $H^{(2)}$ and $H^{(4)}$ are all proportional to the identity matrix $\sigma_0$, therefore only $H^{(1)}$, $H^{(3)}$ and $H^{(5)}$ have to be taken into account.
The first-order correction $H^{(1)}$: 
\begin{equation}
    H^{(1)}=\frac{\Tilde{B}}{2}\sigma_z. 
\end{equation}
The third-order correction: 
\begin{equation}
    H^{(3)}=h_x^{(3)}\sigma_x+h_y^{(3)}\sigma_y+h_z^{(3)}\sigma_z, 
\end{equation}
where the coefficients are: 

\begin{widetext}
\begin{subequations}
\begin{align}
    h_x^{(3)} &=\frac{\Tilde{\alpha}\cos{\phi}\sin{\phi}}{\hbar\omega_0^2}(\Tilde{E}_\mathrm{ac,1}\omega_1 \cos{\omega_1 t}+\Tilde{E}_\mathrm{ac,2}\omega_2 \cos{\omega_2 t})+\frac{\Tilde{B}\Tilde{\alpha}\cos^2{\phi}
    }{\hbar^2 \omega_0^2}(\Tilde{E}_\mathrm{ac,1}\sin{\omega_1 t}+\Tilde{E}_\mathrm{ac,2}\sin{\omega_2 t}) , \label{eq:bi_oneop_Heff31}\\
    h_y^{(3)} &=-\frac{\Tilde{\alpha}\cos^2{\phi}}{\hbar\omega_0^2}(\Tilde{E}_\mathrm{ac,1}\omega_1 \cos{\omega_1 t}+\Tilde{E}_\mathrm{ac,2}\omega_2 \cos{\omega_2 t})+ \frac{\Tilde{B}\Tilde{\alpha}\cos{\phi}\sin{\phi}
    }{\hbar^2 \omega_0^2}(\Tilde{E}_\mathrm{ac,1}\sin{\omega_1 t}+\Tilde{E}_\mathrm{ac,2}\sin{\omega_2 t}),\label{eq:bi_oneop_Heff32}\\
      h_z^{(3)}&=-\frac{\Tilde{B}\Tilde{\alpha}^2}{\hbar^2 \omega_0^2}-\frac{\Tilde{\alpha} \sin{\phi}}{\hbar\omega_0^2}(\Tilde{E}_\mathrm{ac,1}\omega_1 \cos{\omega_1 t}+\Tilde{E}_\mathrm{ac,2}\omega_2 \cos{\omega_2 t}).\label{eq:bi_oneop_Heff33}
\end{align}
\end{subequations}
\end{widetext}

The fifth-order correction $H^{(5)}$: 
\begin{equation}
    H^{(5)}=h_x^{(5)}\sigma_x+h_y^{(5)}\sigma_y+h_z^{(5)}\sigma_z, 
\end{equation}

Where the coefficients are: 

\allowdisplaybreaks

\begin{widetext}
\begin{subequations}
    \begin{align}
        h_x^{(5)} &= \frac{\Tilde{E}_\mathrm{ac,1}\Tilde{\alpha}\omega_1(\hbar^2\omega_1^2-3\Tilde{\alpha}^2)\sin{2\phi}}{2\hbar^3 \omega_0^4}\cos{\omega_1 t}+\frac{\Tilde{E}_\mathrm{ac,2}\Tilde{\alpha}\omega_2 (\hbar^2\omega_2^2-3\Tilde{\alpha}^2)\sin{2\phi}}{2\hbar^3\omega_0^4}\cos{\omega_2 t}+\notag \\
          &\quad +\frac{\Tilde{B}\Tilde{\alpha}(\Tilde{B}^2-3\Tilde{\alpha}^2)\cos^2{\phi}}{\hbar^4\omega_0^4}(\Tilde{E}_\mathrm{ac,1}\sin{\omega_1 t}+\Tilde{E}_\mathrm{ac,2}\sin{\omega_2 t})-\frac{\Tilde{B}\Tilde{\alpha}^2\cos{\phi}\sin^2{\phi}}{\hbar^4 \omega_0^4}(\Tilde{E}_\mathrm{ac,1}\sin{\omega_1 t}+\Tilde{E}_\mathrm{ac,2}\sin{\omega_2 t})^2, \label{eq:bichro_oneop_Heff51} \\
         h_y^{(5)}&=\frac{\Tilde{E}_\mathrm{ac,1}\Tilde{\alpha}\omega_1(3\Tilde{\alpha}^2-\hbar^2\omega_1^2)\cos^2{\phi}}{\hbar^3 \omega_0^4}\cos{\omega_1 t}+ \frac{\Tilde{E}_\mathrm{ac,2}\Tilde{\alpha}\omega_2(3\Tilde{\alpha}^2-\hbar^2\omega_2^2)\cos^2{\phi}}{\hbar^3 \omega_0^4}\cos{\omega_2 t}+\notag \\
         &\quad +\frac{\Tilde{B}\Tilde{\alpha}(\Tilde{B}^2-3\Tilde{\alpha}^2)\cos{\phi}\sin{\phi}}{\hbar^4\omega_0^4}(\Tilde{E}_\mathrm{ac,1}\sin{\omega_1 t}+\Tilde{E}_\mathrm{ac,2}\sin{\omega_2 t})+\frac{\Tilde{B}\Tilde{\alpha}^2\cos^2{\phi}\sin{\phi}}{\hbar^4\omega_0^4}(\Tilde{E}_\mathrm{ac,1}\sin{\omega_1 t}+\Tilde{E}_\mathrm{ac,2}\sin{\omega_2 t})^2,
         \label{eq:bichro_oneop_Heff52} \\
         h_z^{(5)}&=-\frac{\Tilde{B}\Tilde{\alpha}^2(4\Tilde{B}^2+\Tilde{E}_\mathrm{ac,1}^2+\Tilde{E}_\mathrm{ac,2}^2-16\Tilde{\alpha}^2)}{4\hbar^4\omega_0^4}-\frac{\Tilde{B}(\Tilde{E}_\mathrm{ac,1}^2+\Tilde{E}_\mathrm{ac,2}^2)\Tilde{\alpha}^2\cos{2\phi}}{4\hbar^4\omega_0^4}+\frac{\Tilde{B}\Tilde{E}_\mathrm{ac,1}^2\Tilde{\alpha}^2\cos^2{\phi}}{2\hbar^4\omega_0^4}\cos{(2\omega_1 t)}+ \notag \\ &\quad +\frac{\Tilde{B}\Tilde{E}_\mathrm{ac,2}^2\Tilde{\alpha}^2\cos^2{\phi}}{2\hbar^4\omega_0^4}\cos{(2\omega_2 t)}+\frac{\Tilde{E}_\mathrm{ac,1}\Tilde{\alpha}\omega_1(3\Tilde{\alpha}^2-\hbar^2\omega_1^2)\sin{\phi}}{\hbar^3\omega_0^4}\cos{\omega_1 t}+ \notag \\ & \quad +\frac{\Tilde{E}_\mathrm{ac,2}\Tilde{\alpha}\omega_2(3\Tilde{\alpha}^2-\hbar^2\omega_2^2)\sin{\phi}}{\hbar^3\omega_0^4}\cos{\omega_2 t}-\frac{2\Tilde{B}\Tilde{E}_\mathrm{ac,1}\Tilde{E}_\mathrm{ac,2}\Tilde{\alpha}^2\cos^2{\phi}}{\hbar^4\omega_0^4}\sin{\omega_1 t}\sin{\omega_2 t} .      
  \label{eq:bichro_oneop_Heff53}
    \end{align}
\end{subequations}
\end{widetext}

\begin{figure}[h]
\centering
\includegraphics[width=\columnwidth]{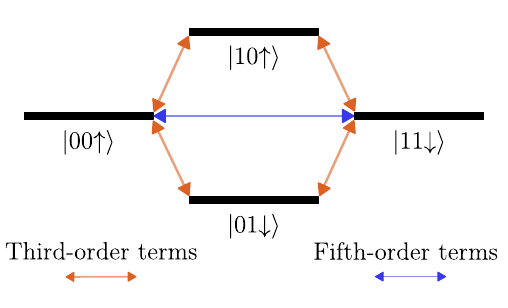}

\caption {\label{fig:Floquet} \footnotesize{\textbf{Fifth-order contributions to the bichromatic process.} The two relevant Floquet levels to the bichromatic driving are $\ket{00{\uparrow}}$ and $\ket{11{\downarrow}}$, where the first two numbers are the photon numbers corresponding to the two driving frequencies. A direct coupling between these states is possible via a fifth-order effective matrix element (blue arrow), and third-order matrix elements participate in second-order couplings. In this figure only two intermediate Floquet levels are shown, $\ket{10{\uparrow}}$ and $\ket{01{\downarrow}}$.}}
\end{figure}

Using this effective Hamiltonian up to the fifth order, the Floquet matrix can be constructed. To describe the bichromatic process, we use the generalization of Floquet theory, the two-mode Floquet theory \cite{ho1983semiclassical, gyorgy2022electrically}. The two Floquet states relevant to the dynamics are $\ket{n_1=0, n_2=0,\uparrow}$ and $\ket{n_1=1, n_2=1,\downarrow}$, where $n_1$ and $n_2$ denote the Fourier-indices (or photon numbers) corresponding to driving frequencies $\omega_1$ and $\omega_2$. Note that in the fifth-order correction of the effective Hamiltonian, Eq.~\eqref{eq:bichro_oneop_Heff51} and Eq.~\eqref{eq:bichro_oneop_Heff52}, terms appear which are proportional to $\sin{\omega_1 t}\sin{\omega_2 t}$, which means that a direct, fifth-order matrix element will couple the two Floquet-levels, and therefore contribute to the Rabi frequency. Fifth-order contributions can also come from the third-order corrections of the effective Hamiltonian, Eq.~\eqref{eq:bi_oneop_Heff31} and Eq.~\eqref{eq:bi_oneop_Heff32}, which do not give direct matrix elements between the two Floquet-states, but still contribute through second-order perturbation theory. These different contributions are represented in Fig. \ref{fig:Floquet}. We have to take the Floquet-matrix and derive an effective $2\times 2$ Floquet-matrix containing the relevant Floquet-states by using second-order time-independent Schrieffer-Wolff transformation. This way all fifth-order contributions are captured, and the higher-order corrections can be neglected.  

This procedure yields an effective Floquet-matrix with the form of Eq.~\eqref{eq:HFeff}, from which the splitting frequency (Eq.~\eqref{eq:split}), the Rabi frequency (Eq.~\eqref{eq:oneop_Rabi}) and the Bloch-Siegert shift (Eq.~\eqref{eq:BS1main}) can be read off.

\subsection{Bichromatic driving with two gates}
Bichromatic driving can also be realized by two different driving gates, which create oscillating electric fields along the $x$ and $y$ direction, for the form of the electric driving term $H_E$ see Eq.~\eqref{eq:electric_twoop}. 
We study the same bichromatic process with resonance condition Eq.~\eqref{eq:resonance}, again a fifth-order TDSW is required to describe the dynamics, see Eq.~\eqref{eq:Heff_fifth}. 
The even order terms, $H^{(0)}$, $H^{(2)}$ and $H^{(4)}$ are proportional to the identity $\sigma_0$, the first-order $H^{(1)}$: 
\begin{equation}
    H^{(1)}=\frac{\Tilde{B}}{2}\sigma_z. 
\end{equation}
The third-order correction: 
\begin{equation}
    H^{(3)}=h_x^{(3)}\sigma_x+h_y^{(3)}\sigma_y+h_z^{(3)}\sigma_z, 
\end{equation}
where the three coefficients are: 

\allowdisplaybreaks
\begin{widetext}
\begin{subequations}
\begin{align}
    h_x^{(3)} &=\frac{\Tilde{E}_\mathrm{ac,2}\Tilde{\alpha}\omega_2 \sin^2{\phi}}{\hbar\omega_0^2}\cos{\omega_2 t}+\frac{\Tilde{E}_\mathrm{ac,1}\Tilde{\alpha}\omega_1 \sin{2\phi}}{2\hbar\omega_0^2}\cos{\omega_1 t}+\frac{\Tilde{B}\Tilde{E}_\mathrm{ac,1}\Tilde{\alpha}\cos^2{\phi}}{\hbar^2\omega_0^2}\sin{\omega_1 t}+\frac{\Tilde{B}\Tilde{E}_\mathrm{ac,2}\Tilde{\alpha}\sin{2\phi}}{2\hbar^2\omega_0^2}\sin{\omega_2 t} , \label{eq:bi_twoop_Heff31}\\
    h_y^{(3)} &=-\frac{\Tilde{E}_\mathrm{ac,1}\Tilde{\alpha}\omega_1 \cos^2{\phi}}{\hbar\omega_0^2}\cos{\omega_1 t}-\frac{\Tilde{E}_\mathrm{ac,2}\Tilde{\alpha}\omega_2 \sin{2\phi}}{2\hbar\omega_0^2}\cos{\omega_2 t}+\frac{\Tilde{B}\Tilde{E}_\mathrm{ac,1}\Tilde{\alpha}\sin{2\phi}}{2\hbar^2\omega_0^2}\sin{\omega_1 t}+\frac{\Tilde{B}\Tilde{E}_\mathrm{ac,2}\Tilde{\alpha}\sin^2{\phi}}{\hbar^2\omega_0^2}\sin{\omega_2 t},\label{eq:bi_twoop_Heff32}\\
      h_z^{(3)}&=-\frac{\Tilde{B}\Tilde{\alpha}^2}{\hbar^2\omega_0^2}+\frac{\Tilde{E}_\mathrm{ac,2}\Tilde{\alpha}\omega_2 \cos{\phi}}{\hbar\omega_0^2}\cos{\omega_2 t}-\frac{\Tilde{E}_\mathrm{ac,1}\Tilde{\alpha}\omega_1 \sin{\phi}}{\hbar\omega_0^2}\cos{\omega_1 t} .\label{eq:bi_twoop_Heff33}
\end{align}
\end{subequations}
\end{widetext}
The fifth-order correction: 
\begin{equation}
    H^{(5)}=h_x^{(5)}\sigma_x+h_y^{(5)}\sigma_y+h_z^{(5)}\sigma_z,
\end{equation}
the constants are 

\begin{widetext}
\allowdisplaybreaks
\begin{subequations}
\begin{align}
    h_x^{(5)} &= -\frac{\Tilde{B}\Tilde{E}_\mathrm{ac,1}^2\Tilde{\alpha}^2\cos{\phi}\sin^2{\phi}}{2\hbar^4\omega_0^4}+\frac{\Tilde{B}\Tilde{E}_\mathrm{ac,2}^2\Tilde{\alpha}^2\cos{\phi}\sin^2{\phi}}{2\hbar^4\omega_0^4}+\frac{\Tilde{B}\Tilde{E}_\mathrm{ac,1}^2\Tilde{\alpha}^2\cos{\phi}\sin^2{\phi}}{2\hbar^4\omega_0^4}\cos{(2 \omega_1 t)} 
+ \notag \\ & \quad -\frac{\Tilde{B}\Tilde{E}_\mathrm{ac,2}^2\Tilde{\alpha}^2\cos{\phi}\sin^2{\phi}}{2\hbar^4\omega_0^4}\cos{(2\omega_2 t)}+\frac{\Tilde{E}_\mathrm{ac,2}\Tilde{\alpha}\omega_2(\hbar^2\omega_2^2-3\Tilde{\alpha}^2)\sin^2{\phi}}{\hbar^3\omega_0^4}\cos{\omega_2 t}+ \notag \\ & \quad +\frac{\Tilde{E}_\mathrm{ac,1}\Tilde{\alpha}\omega_1(\hbar^2\omega_1^2-3\Tilde{\alpha}^2)\sin{\phi}\cos{\phi}}{\hbar^3\omega_0^4}\cos{\omega_1 t}+\frac{\Tilde{B}\Tilde{E}_\mathrm{ac,1}\Tilde{\alpha}(\Tilde{B}^2-3\Tilde{\alpha}^2)\cos^2{\phi}}{\hbar^4\omega_0^4}\sin{\omega_1 t} + \notag \\ & \quad +\frac{3\Tilde{E}_\mathrm{ac,1}\Tilde{E}_\mathrm{ac_2}\Tilde{\alpha}^2\omega_2\cos{\phi}}{\hbar^3\omega_0^4}\sin{\omega_1 t}\cos{\omega_2 t}-\frac{3\Tilde{E}_\mathrm{ac,1}\Tilde{E}_\mathrm{ac,2}\Tilde{\alpha}^2\omega_1 \cos{\phi}}{\hbar^3\omega_0^4}\sin{\omega_2 t}\cos{\omega_1 t}+ \notag \\ & \quad +\frac{\Tilde{B}\Tilde{E}_\mathrm{ac,2}\Tilde{\alpha}(\Tilde{B}^2-3\Tilde{\alpha}^2)\cos{\phi}\sin{\phi}}{\hbar^4\omega_0^4}\sin{\omega_2 t}+\frac{\Tilde{B}\Tilde{\alpha}^2\Tilde{E}_\mathrm{ac,1}\Tilde{E}_\mathrm{ac,2}}{2\hbar^4\omega_0^4}(\sin{3\phi}-\sin{\phi})\sin{\omega_1 t}\sin{\omega_2 t} , \label{eq:bi_twoop_Heff51}\\
    h_y^{(5)} &=\frac{\Tilde{E}_\mathrm{ac,1}\Tilde{\alpha}\omega_1(3\Tilde{\alpha}^2-\hbar^2\omega_1^2)\cos^2{\phi}}{\hbar^3\omega_0^4}\cos{\omega_1 t}+\frac{\Tilde{B}\Tilde{E}_\mathrm{ac,1}^2\Tilde{\alpha}^2\cos^2{\phi}\sin{\phi}}{2\hbar^4\omega_0^4}-\frac{\Tilde{B}\Tilde{E}_\mathrm{ac,2}^2\Tilde{\alpha}^2\cos^2{\phi}\sin{\phi}}{2\hbar^4\omega_0^4}+ \notag \\ & \quad -\frac{\Tilde{B}\Tilde{E}_\mathrm{ac,1}^2\Tilde{\alpha}^2\cos^2{\phi}\sin{\phi}}{2\hbar^4\omega_0^4}\cos{(2\omega_1 t)}+\frac{\Tilde{E}_\mathrm{ac,2}\Tilde{\alpha}\omega_2(3\Tilde{\alpha}^2-\hbar^2\omega_2^2)\cos{\phi}\sin{\phi}}{\hbar^3\omega_0^4}\cos{\omega_2 t}+ \notag \\ &\quad +\frac{\Tilde{B}\Tilde{E}_\mathrm{ac,2}^2\Tilde{\alpha}^2\cos^2{\phi}\sin{\phi}}{2\hbar^4\omega_0^4}\cos{(2\omega_2 t)} +\frac{\Tilde{B}\Tilde{E}_\mathrm{ac,1}\Tilde{\alpha}(\Tilde{B}^2-3\Tilde{\alpha}^2)\sin{\phi}\cos{\phi}}{\hbar^4\omega_0^4}\sin{\omega_1 t}+\notag \\ & \quad +\frac{3\Tilde{E}_\mathrm{ac,1}\Tilde{E}_\mathrm{ac,2}\Tilde{\alpha}^2\omega_2 \sin{\phi}}{\hbar^3\omega_0^4}\sin{\omega_1 t}\cos{\omega_2 t}-\frac{3\Tilde{E}_\mathrm{ac,1}\Tilde{E}_\mathrm{ac,2}\Tilde{\alpha}^2\omega_1\sin{\phi}}{\hbar^3\omega_0^4}\sin{\omega_2 t}\cos{\omega_1 t}+\notag \\ & \quad +\frac{\Tilde{B}\Tilde{E}_\mathrm{ac,2}\Tilde{\alpha}(\Tilde{B}^2-3\Tilde{\alpha}^2)\sin^2{\phi}}{\hbar^4\omega_0^4}\sin{\omega_2 t}-\frac{\Tilde{B}\Tilde{E}_\mathrm{ac,1}\Tilde{E}_\mathrm{ac,2}\Tilde{\alpha}^2(\cos{\phi}+\cos{3\phi})}{2\hbar^4\omega_0^4}\sin{\omega_1 t}\sin{\omega_2 t},\label{eq:bi_twoop_Heff52}\\
      h_z^{(5)}&=-\frac{\Tilde{B}\Tilde{\alpha}^2(4\Tilde{B}^2+\Tilde{E}_\mathrm{ac,1}^2+\Tilde{E}_\mathrm{ac,2}^2-16\Tilde{\alpha}^2+\Tilde{E}_\mathrm{ac,1}^2\cos{2\phi}-\Tilde{E}_\mathrm{ac,2}^2\cos{2\phi})}{4\hbar^4\omega_0^4}+\frac{\Tilde{B}\Tilde{E}_\mathrm{ac,1}^2\Tilde{\alpha}^2\cos^2{\phi}}{2\hbar^4\omega_0^4}\cos{2\omega_1 t}+\notag \\ & \quad +\frac{\Tilde{E}_\mathrm{ac,2}\Tilde{\alpha}\omega_2(\hbar^2\omega_2^2-3\Tilde{\alpha}^2)\cos{\phi}}{\hbar^3\omega_0^4}\cos{\omega_2 t}+\frac{\Tilde{E}_\mathrm{ac,1}\Tilde{\alpha}\omega_1(3\Tilde{\alpha}^2-\hbar^2\omega_1^2)\sin{\phi}}{\hbar^3\omega_0^4}\cos{\omega_1 t}+\notag \\ & \quad +\frac{\Tilde{B}\Tilde{E}_\mathrm{ac,2}^2\Tilde{\alpha}^2\sin^2{\phi}}{2\hbar^4\omega_0^4}\cos{2\omega_2 t}-\frac{\Tilde{B}\Tilde{E}_\mathrm{ac,1}\Tilde{E}_\mathrm{ac,2}\Tilde{\alpha}^2\sin{2\phi}}{\hbar^4\omega_0^4}\sin{\omega_1 t}\sin{\omega_2 t}.\label{eq:bi_twoop_Heff53}
\end{align}
\end{subequations}
\end{widetext}
Similarly to the case of bichromatic driving using a single gate, based on the effective Hamiltonian the Floquet-matrix can be constructed. We apply second-order Schrieffer-Wolff transformation to derive an effective $2\times 2$ Floquet-matrix, which yields the splitting frequency (Eq.~\eqref{eq:split}), the Rabi frequency (Eq.~\eqref{eq:twoop_Rabi}) and the Bloch-Siegert shift (Eq.~\eqref{eq:BS2main}). 

\bibliography{paper}
\end{document}